\documentclass[12pt]{iopart}

\usepackage{graphicx,iopams}
\newcommand{\llangle}{\left\langle}
\newcommand{\rrangle}{\right\rangle}

\begin{document}

\title{Dynamics of confined suspensions of swimming particles}
\author{Juan P Hernandez-Ortiz$^{1}$, Patrick T Underhill$^2$ and Michael D Graham$^2$}

\address{$^1$Departamento de Materiales, Universidad Nacional de Colombia
              Sede Medell\'in, Carrera 80 \# 65-223, Bloque M3-050,
              Medell\'in, Colombia}
\address{$^2$Department of Chemical and Biological Engineering,
              University of Wisconsin-Madison, Madison, WI 53706-1691 }
\ead{graham@engr.wisc.edu}

\date{\today}

\begin{abstract}
Low Reynolds number direct simulations of large populations of hydrodynamically interacting
swimming particles confined between planar walls are performed. The results of simulations are
compared with a theory that describes dilute suspensions of swimmers. The theory yields scalings
with concentration for diffusivities and velocity fluctuations as well as a prediction of the
fluid velocity spatial autocorrelation function. Even for uncorrelated swimmers, the theory
predicts anticorrelations between nearby fluid elements that correspond to vortex-like swirling
motions in the fluid with length scale set by the size of a swimmer and the slit height. Very
similar results arise from the full simulations indicating either that correlated motion of the
swimmers is not significant at the concentrations considered or that the fluid phase
autocorrelation is not a sensitive measure of the correlated motion. This result is in stark
contrast with results from unconfined systems, for which the fluid autocorrelation captures
large-scale collective fluid structures. The additional length scale (screening length) introduced
by the confinement seems to prevent these large-scale structures from forming.
\end{abstract}

\submitto{\JPCM}

\maketitle

\section{Introduction}\label{sec:intro}

Experimental observations of suspensions of swimming microorganisms illustrate a number of
fascinating phenomena that are still poorly understood. Correlations of the swimmers result in
jets and swirling motions on scales larger than that of a single
organism~\cite{MendelsonWhirls99,GoldsteinKesslerSPP04}. The organisms also form local nematic
ordering though they have no global nematic behavior~\cite{KesslerZoomingNematicExpFluids2007}. The
collective behavior leads to swimmer velocities larger than that of an isolated
organism~\cite{AransonSwimExpts07} and enhanced transport in the fluid~\cite{WuLibchaber00}.

There are many open questions regarding these important observations both in terms of what leads
to the phenomena and their biological significance. Some models consider ``local'' interactions
between nearby organisms within a finite range, and how those can lead to large-scale collective
behavior. These interactions may take the form of ad-hoc rules~\cite{VicsekSPP95} or direct steric
interactions between the swimmers~\cite{MarchettiHardRodsPRE08}. Other models consider long-ranged
hydrodynamic interactions between swimmers, which decay as $r^{-2}$ in an unbounded
domain~\cite{hernandez-spp,ShelleyRodsPRL07,PedleyDiffusionJFM2007,NottJFM2008,underhill-spp}. The
relative importance of these phenomena is still not fully understood.

Many experiments observing these phenomena have been performed in droplets or thin films, but the
influence of confinement on the observations is not clear. Confinement sterically hinders the
organisms and also affects the hydrodynamic interactions, both between the organisms and between
the organism and the boundary. The boundaries may also play a role in transport of nutrients,
which may, in turn, affect motion of the organisms. For example, oxygen levels may be different
near surfaces, altering the motion of the organisms. For droplets or films, the rigidity of the
surface from secreted molecules may affect the fluid boundary condition and dynamics of the
organisms~\cite{AransonSwimExpts07,KesslerZoomingNematicExpFluids2007}.

Most experiments mentioned above have concentrated on two types of bacteria, \emph{E.~coli} and
\emph{B.~subtilis}. Both organisms use flagella to propel themselves forward through the fluid.
However, other types of organisms propel through a fluid by other mechanisms, such as pulling
themselves forward from the front~\cite{BrayBook01}. What role the propulsion method plays on the
collective behavior has not been clarified, although computational and theoretical evidence suggests that organisms pushed from behind display more complex collective behavior than those pulled from the front \cite{Saintillan:2008p45,underhill-spp}. In general, what biological significance the above
phenomena play in the function of the microorganisms remains unclear.

In this article we focus on the role that confinement plays in the collective behavior. The only
prior publication using computational models to investigate how hydrodynamic interactions affect
the collective behavior of swimming microorganisms in confined environments is by Hern\'andez
\emph{et al}.~\cite{hernandez-spp}. They showed that hydrodynamic interactions were sufficient to
produce many qualitative phenomena seen in experiments, including increased transport and swirls
in the fluid. We focus on confined systems not only because many experiments have been done in
confined environments, but also to draw comparisons with unconfined dynamics. Simulations in three-dimensional periodic domains have shown changes in dynamics with the size of
the domain~\cite{underhill-spp} that are thought to arise because of long-range structures that
fill the entire domain. Some experiments have attempted to measure unconfined dynamics by
examining the response far away from the confining
walls~\cite{Shivashankar03,Shivashankar04,KochPairQuadrupolePoF2007}. However, because of the
long-ranged nature of hydrodynamics, the point at which confined systems with very large gaps
reduces to an unconfined system is not obvious.

We examine the behavior at different levels of confinement and at different concentrations ranging
from the dilute limit into the semidilute regime. However the concentration is still low enough
that we do not expect the steric interactions between the organisms to be the dominant interaction
and lead to nematic-like structures. Instead we expect the long-ranged hydrodynamic interactions
to play an important role. The confinement boundaries alter these hydrodynamic interactions. The
walls also induce a non-uniform concentration profile within the domain, with the organisms
concentrating at the walls. Finally, the walls introduce a new length scale which screens
hydrodynamics and alters flow structures over larger length scales. We will show in this article
how the dynamics depend on confinement and how they compare with theoretical predictions.

\section{Swimmer Model}\label{sec:model}
Consider a suspension of $N$ neutrally buoyant rodlike swimmers confined between two planar walls.
Directions $x_1$ and $x_2$ are periodic of side length $L$, and the walls are separated in the
$x_3$-direction by a distance $2H$. Each swimmer has a characteristic length $\ell$, a
characteristic width $w$, and in isolation would move in a straight line with a speed
$v_{\mathrm{is}}$. It is assumed that the Reynolds number, $Re = v_{\mathrm{is}} H / \nu \ll 1 $,
where $\nu$ is the fluid kinematic viscosity, in which case the fluid motion is governed by the
Stokes equation. To allow treatment of large populations ($> 10^3$ swimmers) over long times, a
very simple model of each swimmer is adopted, following our previous
work~\cite{hernandez-spp,underhill-spp}.

Each self-propelled particle is modeled as two beads connected by a stiff spring with equilibrium
length $\ell$ as shown in figure~1. The unit vector pointing along the swimmer axis from bead 1 to
bead 2 is denoted $\mathbf{n}$. Propulsion is provided by a ``phantom flagellum'' that we do not
treat explicitly, but only through its effect on the swimmer body and the fluid. The bead which is
connected to the flagellum, denoted bead 1, feels a force $\mathbf{f}^{\mathrm{f}}$ exerted on it.
However, the flagellum also exerts a force $- \mathbf{f}^{\mathrm{f}}$ on the fluid. This force on
the fluid occurs at the position of bead 1. With this model we can consider ``pushers'' or
``pullers'' depending on whether $\mathbf{f}^{\mathrm{f}}$ is parallel or anti-parallel to
$\mathbf{n}$, respectively. A pusher sends fluid away from it fore and aft, with fluid moving
toward its ``waist'', and vice versa for a puller. Whether a real swimmer is a pusher or a puller
depends on the specific mechanism of locomotion (cf.~\cite{BrayBook01})-- a cell whose flagella
propel it forward predominantly from behind would be a pusher~\cite{hernandez-spp,underhill-spp}.
We will focus primarily on pushers in this article, though some results with pullers are
presented.

To measure the concentration of the swimmers we define an effective volume fraction
$\phi_{\mathrm{e}} = \pi N \ell^3 / (12 L^2 H)$ -- this would be the true volume fraction if the
swimmers were spheres of diameter $\ell$. In the present geometry, swimmers in dilute
systems form layers near the two walls. Therefore, we find it convenient to also define an effective
area fraction if all swimmers resided in the layers. It is defined as $\psi_{\mathrm{e}} = C \pi N
\ell^2 / (4L^2)$, where the constant $C=1/2$ if the gap is large enough to allow for a layer at
each wall ($2H > 2 w$), or $C=1$ for extreme confinements, e.g.~monolayers.

\begin{figure}
\begin{center}
\includegraphics[width=0.47\textwidth]{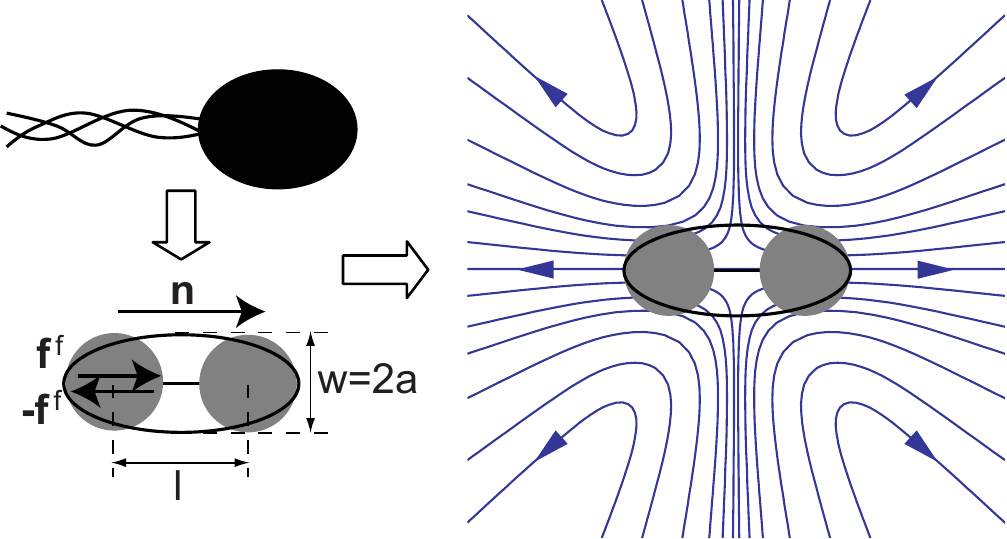}
\caption{Illustration of a pushing organism, our swimmer model, and the fluid disturbance they
cause in an unbounded domain. Our model shows the hydrodynamic radius of the beads and the
ellipsoidal excluded volume. The double-arrow signifies the phantom flagellum force acting on the
bead and an opposite force acting on the fluid. Both forces act at the center of the first bead.
The blue lines represent streamlines of the axisymmetric fluid disturbance. A puller would produce
the same streamlines, but with the direction of the velocities reversed. }
\end{center}
\label{fig:swimmer}
\end{figure}

The force in the ``stiff'' connector between the beads follows a finitely extensible non-linear
elastic (FENE) spring model with a non-zero equilibrium length equal to the swimmer size
$\ell$~\cite{dpl2}:
\begin{equation}
\mathbf{f}_{\nu,k}^{\mathrm{c}} = h \mathbf{Q}_{\nu} \frac{1-\ell/| \mathbf{Q}_{\nu} | }{1-[ (|
\mathbf{Q}_{\nu} | - \ell )/( \ell_{\mathrm{m}}-\ell ) ]^2}, \label{eq:fene}
\end{equation}
where $h$ is the spring constant, $\mathbf{Q}_{\nu}$ is the connector vector from bead 1 to bead 2
on swimmer $\nu$, and $\ell_{\mathrm{m}}$ is the swimmer maximum size. With this spring law, the
swimmer will shrink slightly for a pushing flagellum, or expand for a pulling one. The
values of $h$ and $\ell_{\mathrm{m}}$ are chosen so that the spring approximates a rigid
constraint while still allowing timesteps that are not prohibitively small. For the results
presented here, $ h = \mathrm{f}^{\mathrm{f}} /(0.1 \ell) $ and $ \ell_{\mathrm{m}} = 1.15 \ell $.
The swimmers also interact through an excluded volume potential, which is taken as the repulsive
portion of the Gay-Berne potential~\cite{AllenGermanoMolPhys06}; this potential is widely used in
molecular simulations to model steric repulsions between rodlike objects. The size and aspect
ratio of the excluded volume potential to related to the size of the swimmer $\ell$ and the
hydrodynamic radius of a bead $a$ as illustrated in figure~\ref{fig:swimmer}. The width of the
excluded volume is taken as $w=2 a$ and the length as $\ell + 2a$. This means that the excluded
volume restricts bead positions such that the hydrodynamic radii do not overlap. It also gives an aspect
ratio of $\gamma=1+\ell/(2a)$. The model presented here has $\ell=3 a$, and therefore an aspect
ratio of $\gamma=2.5$.

The motion of the swimmers is determined by the force balance (neglecting inertia because of the
small size of a microorganism) for each bead ($k=1,2$) of a swimmer $\nu$, as follows
\begin{equation}
\mathbf{f}_{\nu,k}^{\mathrm{h}} + \mathbf{f}_{\nu,k}^{\mathrm{c}} +
\mathbf{f}_{\nu,k}^{\mathrm{x}} + \delta_{k1} \mathbf{f}_\nu^{\mathrm{f}} = 0 \qquad \textrm{for}
\qquad k=1,2, \label{eq:force-bead}
\end{equation}
where $\delta_{ij}$ is the Kronecker delta, $\mathbf{f}_{\nu,k}^{\mathrm{h}}$ is the hydrodynamic
drag force, $\mathbf{f}_{\nu,k}^{\mathrm{c}}$ is the connector (spring) force and
$\mathbf{f}_{\nu,k}^{\mathrm{x}}$ are the bead-bead, bead-swimmer and bead-wall excluded volume
forces. Notice that the force balance for bead 1 differs from the balance on bead 2 by the
presence of the flagellum force, $\mathbf{f}_\nu^{\mathrm{f}}$.

In our simulations, each bead will be treated as a point particle. In this situation, the hydrodynamic drag force on a bead $k$ of swimmer $\nu$ is given by a generalization of Stokes'
law~\cite{Jendrejack03b}:
\begin{equation}
\mathbf{f}_{\nu,k}^{\mathrm{h}} = -\zeta\left( {\bf v}_{\nu,k} - {\bf u}_{\nu,k} \right),
\label{eq:stokes-law}
\end{equation}
where $\zeta=6\pi\eta a$ is the Stokes drag coefficient on a bead with hydrodynamic radius $a$ in
a fluid with viscosity $\eta$, ${\bf v}_{\nu,k}=\dot{\bf x}_{\nu,k}$ is the velocity of the bead,
where ${\bf x}_{\nu,k}$ is the position (cartesian coordinates) of the bead, and ${\bf u}_{\nu,k}$
is the fluid velocity at the bead position. There are two contributions to this fluid velocity: the first is the motion driven by the forces exerted by the other beads (and flagella) in the system, and the second is the correction to the velocity experienced by the bead due to the presence of confining walls. This correction leads for example to the decrease in bead mobility found in a confining geometry. Both of these contributions are determined simultaneously by the methodology used here.  Equations~(\ref{eq:stokes-law}) and
(\ref{eq:force-bead}) give an evolution equation for the bead positions as follows,
\begin{equation}
\frac{\rmd {\bf x}_{\nu,k}}{\rmd t} = {\bf u}_{\nu,k}
  + \frac{1}{\zeta}\left( {\bf f}_{\nu,k}^{\mathrm{c}} + {\bf f}_{\nu,k}^{\mathrm{x}}
  + \delta_{1k}{\bf f}_{\nu}^{\mathrm{f}} \right).
\label{eq:evolution}
\end{equation}

For $Re=0$, the fluid velocity at a point ${\bf x}$ due to a collection of point-forces is
calculated by summing over all forces times the Green's function $\mathbf{G}$ for the Stokes
equation for the geometry and boundary conditions of interest. Considering the beads of each of
the $N$ swimmers as point-forces and including the disturbance due to the phantom flagellum, we
write the fluid velocity at a point ${\bf x}$ as
\begin{equation}
{\bf u}({\bf x}) = \sum_{\mu=1}^{N}\sum_{l=1}^2 {\bf G}(\mathbf{x},\mathbf{x}_{\mu,l})
\cdot\left(- {\bf f}_{\mu,l}^{\mathrm{h}} - \delta_{l1}{\bf f}_\mu^{\mathrm{f}} \right) ,
\end{equation}
where the first term represents the direct hydrodynamic forces exerted by the beads on the fluid
and the second term represents the disturbance caused by the phantom flagella. These two
contributions can be combined using the force balance on each bead to become
\begin{equation}\label{eq:ufluid_elimflagella}
{\bf u}({\bf x}) = \sum_{\mu=1}^{N}\sum_{l=1}^2 {\bf G}({\bf x},{\bf x}_{\mu,l}) \cdot\left({\bf
f}_{\mu,l}^{\mathrm{c}}+{\bf f}_{\mu,l}^{\mathrm{x}}\right).
\end{equation}
Similarly, the fluid velocity at the position of a bead $k$ of swimmer $\nu$ is calculated by
excluding the singular part of the fluid velocity generated by the bead:
\begin{equation}
{\bf u}_{\nu,k} = \sum_{\mu=1}^{N}\sum_{l=1}^{2} ({\bf G}({\bf
x}_{\nu,k},{\bf x}_{\mu,l})-\delta_{\nu\mu}\delta_{kl}{\bf G}_\infty({\bf
x}_{\nu,k}-{\bf x}_{\mu,l}))\cdot\left({\bf f}_{\mu,l}^{\mathrm{c}}+{\bf f}_{\mu,l}^{\mathrm{x}}
\right). \label{eq:perturb3}
\end{equation}
Here ${\bf G}_\infty$ is the Oseen-Burgers or Stokeslet tensor, the free-space point-force Green's function for Stokes equation. It is given explicitly as:
\begin{equation}
{\bf G}_\infty({\bf x})=\frac{1}{8\pi\eta r}\left( \bdelta +\frac{{\bf xx}}{r^2} \right)
\end{equation}
with $r=|{\bf x}|$ and $\bdelta$ the identity tensor. In an unbounded domain, ${\bf G}$ and ${\bf G}_\infty$ are identical so the $\nu=\mu$, $k=l$ terms in this expression are zero. In a bounded domain, however, ${\bf G}$ contains a finite correction due to the change in the flow induced by the boundaries; subtracting off ${\bf G}_\infty$ when $\nu=\mu$, $k=l$ reveals this finite correction in the overall expression. Substitution of Eq.~(\ref{eq:perturb3}) into the evolution equation for the bead
positions, Eq.~(\ref{eq:evolution}), results in
\begin{equation}
\frac{\rmd {\bf x}_{\nu,k}}{\rmd t} =
  \frac{1}{\zeta}\left(\delta_{1k}{\bf f}_{\nu}^{\mathrm{f}} \right)
+ \sum_{\mu=1}^{N}\sum_{l=1}^{2} {\bf M}_{(\nu,k)(\mu,l)} \cdot\left({\bf
f}_{\mu,l}^{\mathrm{c}}+{\bf f}_{\mu,l}^{\mathrm{x}} \right).
 \label{eq:evolution3}
\end{equation}
Here ${\bf M}_{(\nu,k)(\mu,l)}$ is a $3\times 3$ tensor which constitutes a block of the
$(3\times 2N) \times (3\times 2N)$ mobility tensor, ${\bf M}$:~\cite{kim_karrila}
\begin{equation}
{\bf M}_{(\nu,l)(\mu,k)}= \delta_{\nu\mu}\delta_{kl}\frac{\bdelta}{\zeta} +
({\bf G}({\bf
x}_{\nu,k},{\bf x}_{\mu,l})-\delta_{\nu\mu}\delta_{kl}{\bf G}_\infty({\bf
x}_{\nu,k}-{\bf x}_{\mu,l})).
 \label{eq:mobility}
\end{equation}
Equation~(\ref{eq:evolution3}) can be written in a compact form by
introducing $3\times 2N$ dimensional vectors containing the coordinates and forces of
all beads as follows:
\begin{equation}
\frac{\rmd}{\rmd t}{\bf R} = \frac{1}{\zeta}{\bf F}^{\mathrm{f}} + {\bf M}\cdot{\bf
F}^{\mathrm{b}}, \label{eq:evolution4}
\end{equation}
where ${\bf R}$ contains the coordinates of all beads, ${\bf F}^{\mathrm{f}}$ is a vector with the
flagellum force, whose components are non-zero for bead 1 of each swimmer, and ${\bf
F}^{\mathrm{b}}$ is a vector containing the total non-hydrodynamic and non-flagellar forces on
each bead.

The net force exerted by an isolated swimmer on the fluid is zero -- to leading order in the far field a neutrally buoyant
swimmer is a force dipole. The present approach captures this universal far-field
behavior while neglecting the near-field corrections to the hydrodynamic interactions between
swimmers, which are dependent on the details of the organism. The validity of this approximation
is supported by recent simulation results~\cite{NottJFM2008}. Finally, we note that the limited
set of results we have obtained with multi-bead rod swimmers is qualitatively consistent with
those for the two-bead swimmers.

The fluid velocity $ {\bf M}\cdot{\bf F}^{\mathrm{b}} $ is calculated using the General
Geometry Ewald-like Method (GGEM) introduced by Hern\'andez-Ortiz {\it et
al.}~\cite{hernandez-ggem}. A brief description of the GGEM starts with considering the Stokes
system of equations for a flow driven by a distribution of $2 N$ point forces,
\begin{equation}
\eqalign{ -\nabla p({\bf x}) + \eta\nabla^2{\bf u}({\bf x}) = -\brho({\bf x}) \cr
 \nabla\cdot{\bf u}({\bf x}) = 0 },
\label{eq:stokes1}
\end{equation}
where $\eta$ is the fluid viscosity and the force density is $\brho({\bf x})=\sum_{\nu=1}^{2N}
{\bf f}_\nu\delta({\bf x-x}_\nu)$. Here ${\bf f}_\nu$ is the force exerted on the fluid at point
${\bf x}_\nu$. The solution of~(\ref{eq:stokes1}), can be written in the form
of~(\ref{eq:ufluid_elimflagella}) and combined into the ${\bf M}\cdot{\bf F}$ product. If computed
explicitly, this product is a matrix-vector operation that requires $O(N^2)$ calculations. GGEM
determines the product implicitly for any geometry (with appropriate boundary conditions) without
performing the matrix-vector manipulations. It starts with the restatement of the force-density
expression in~(\ref{eq:stokes1}), $\brho({\bf x}) = \brho_\mathrm{l}({\bf x}) +
\brho_\mathrm{g}({\bf x})$ using a smoothing function $g$, similar to conventional particle-mesh
methods. By linearity of the Stokes equation, the fluid velocity is written as a sum of two parts,
the solution with each force-density separately. The ``local density''
\begin{equation}
\brho_\mathrm{l}({\bf x}) = \sum_{\nu=1}^{2N} {\bf f}_\nu [ \delta({\bf x-x}_\nu) - g({\bf
x-x}_\nu) ]
\end{equation}
drives a local velocity, ${\bf u}_\mathrm{l}({\bf x})$, which is calculated assuming an unbounded
domain: ${\bf u}_{\mathrm{l}}({\bf x}) = \sum_\nu^N{\bf G}_{\mathrm{l}}({\bf x}-{\bf x}_\nu)
\cdot{\bf f}_\nu$, where ${\bf G}_{\mathrm{l}}({\bf x})$ is composed of the free-space Green's
function ${\bf G}_\infty$ minus a smoothed Stokeslet obtained from the solution of Stokes equations
with the forcing term modified by the smoothing function $g$. For the Stokes equations we found
that a modified Gaussian smoothing function defined by
\begin{equation}
g(r) = (\alpha^3/\pi^{3/2})e^{(-\alpha^2r^2)}( 5/2-\alpha^2r^2 )
\end{equation}
yields a simple expression for ${\bf G}_\mathrm{l}({\bf x})$:
\begin{eqnarray}
\eqalign{ {\bf G}_{\mathrm{l}}({\bf x}) &= \frac{1}{8\pi\eta}\left( \bdelta +\frac{{\bf xx}}{r^2} \right)
\frac{\textrm{erfc}(\alpha r)}{r}  \cr
&- \frac{1}{8\pi\eta} \left( \bdelta-\frac{{\bf xx}}{r^2} \right)
\frac{2\alpha}{\pi^{1/2}}e^{(-\alpha^2r^2)}.} \label{eq:hy7}
\end{eqnarray}
Because ${\bf G}_\mathrm{l}({\bf x})$ decays exponentially on the length scale $\alpha^{-1}$, in
practice the local velocity can be computed, as in normal Ewald methods, by only considering
near-neighbors to each particle $\nu$~\cite{hockney_eastwook_book}.

For the present work, the point-particle approximation is not desired; in particular, as the
concentration of particles increases, the probability that particles will overlap, having
unphysical velocities, increases. To avoid this problem, the bead hydrodynamic radius, $a$, can be
used to define a new smoothed-force density which will give a non-singular velocity. This is
achieved by replacing the Stokeslet by a regularized Stokeslet, using the same modified Gaussian
with $\alpha$ replaced by $\xi$, with $\xi\sim a^{-1}$, yielding
\begin{eqnarray}
\eqalign{ {\bf G}_{\mathrm{l}}^{\mathrm{R}}({\bf x}) &= \frac{1}{8\pi\eta}\left( \bdelta +\frac{{\bf
xx}}{r^2} \right)
\left[ \frac{\textrm{erf}(\xi r)}{r} - \frac{\textrm{erf}(\alpha r)}{r} \right] \cr
&+\frac{1}{8\pi\eta} \left( \bdelta-\frac{{\bf xx}}{r^2} \right) \left(
\frac{2\xi}{\pi^{1/2}}e^{(-\xi^2r^2)}- \frac{2\alpha}{\pi^{1/2}}e^{(-\alpha^2r^2)} \right),}
\label{eq:hy8}
\end{eqnarray}
where the superscript $\mathrm{R}$ stands for regularized force density. For
$\xi^{-1}=3a/\sqrt{\pi}$, the maximum fluid velocity is equal to that of a particle with radius
$a$ and the pair mobility remains positive-definite~\cite{hernandez-ggem,power_book}.

The global velocity, ${\bf u}_\mathrm{g}({\bf x})$, is due to the force distribution
$\brho_\mathrm{g}({\bf x})$, which is given by
\begin{equation}
\brho_\mathrm{g}({\bf x}) = \sum_{\nu=1}^{2N} {\bf f}_\nu g({\bf x-x}_\nu) .
\end{equation}
For a general domain, we find the solution to Stokes' equation numerically, requiring that ${\bf
u}_{\mathrm{l}}+{\bf u}_\mathrm{g}$ satisfy appropriate boundary conditions. At a no-slip boundary
we would require ${\bf u}_{\mathrm{g}}({\bf x}) = -{\bf u}_{\mathrm{l}}({\bf x})$. For problems
with periodic boundary conditions, Fourier techniques can be used to guarantee the periodicity of
the global velocity ${\bf u}_\mathrm{g}$. The periodicity on the local velocity, ${\bf
u}_\mathrm{l}$, is obtained using the minimum image convention. In the present case, the global
contribution is calculated on a mesh with $M=M_1 M_2 M_3$ mesh points, with $M_1$, $M_2$, and
$M_3$ the number of mesh points in the $x_1$, $x_2$, and $x_3$ directions. A Fast Fourier
Transform (FFT) method is used in the two periodic directions ($x_1$ and $x_2$) and a second order
finite difference method (FDM) scheme for the confined direction ($x_3$). The FFT is implemented
using FFTW~\cite{FFTW97,FFTW05}, while the FDM is solved with a regular LU decomposition
routine~\cite{nr_press}.

For GGEM, the results should be independent of $\alpha$ but the computational cost of the local
and global calculations depend on $\alpha$. Therefore, we choose the optimal $\alpha$ which
minimizes the total computational cost. In the global calculation, to have an accurate solution,
the mesh size must be smaller than the scale of the smoothing function, which is $\alpha^{-1}$.
Therefore, $M_{1,2,3} \sim \alpha$. The cost of the each LU decomposition scales as $M_3^{3}$ and
there are $M_1 M_2$ number of them, giving a total global cost that scales as $\alpha^5$. Note
that the cost of each FFT scales as $M_1 M_2 \ln ( M_1 M_2 )$ and there are $M_3$ of them giving a
scaling of $\alpha^3 \ln (\alpha) $. This is small compared to the cost of the LU decomposition.
In the local calculation, the contribution of all pairs must be calculated that lie within a
neighbor list determined by the decay of the local Green's function. The local Green's function
decays over a distance $\alpha^{-1}$, so the number of neighbors for each particle scales as $N
\alpha^{-3}$. The calculation must be performed over all pairs, which is the number of particles
times the number of neighbors per particle, resulting in a local calculation cost scaling of $N^2
\alpha^{-3}$. Minimizing the total (local and global) computational cost with respect to $\alpha$
gives an optimal $\alpha$ that scales as $\alpha_{\mathrm{opt}} \sim N^{1/4}$ and a total cost
that scales as $O(N^{5/4})$. If we had chosen a different, linear, method for the FDM calculation,
the global cost would have scaled as $\alpha^3$, leading to an optimal value of
$\alpha_{\mathrm{opt}} \sim N^{1/3}$ and a total computational cost that scaled as $O(N)$.

For our slit geometry, we used for the periodic directions a number of mesh points $M_{1,2} =
\sqrt{2}\alpha L$ while $M_3 = 4\sqrt{2}\alpha H$ for the confined direction. The previous
analysis determined how the optimal value of $\alpha$ changes with system size $N$. However, to
determine the value of the prefactor, and thus the value of $\alpha$ used in the simulations, the
cost of the global and local contributions must be determined on the computers used for the
computation. For our machines, determining the computational cost led to using
$\alpha_{\mathrm{opt}} = 0.042N^{1/4}$.

\section{Theory for Dilute Suspensions}\label{sec:theory}
As a starting point for understanding the dynamics of confined suspensions of swimmers, we present
here a scaling theory valid in the dilute limit, in which the swimmers act almost independently.
We also show the fluid correlations for a suspension of uncorrelated swimmers. These properties
are an important baseline in order to identify whether or not an observed feature of a swimming
suspension arises from collective dynamics.

\subsection{Diffusivity and velocity scaling}
In the dilute limit, swimmers are mostly to be found in layers near each
wall~\cite{hernandez-spp}. The fundamental reason for this is simple -- a swimmer that is not near
a wall is in all likelyhood oriented toward one wall or the other and will eventually collide with
it. Once it collides with a wall it remains there until a fluctuation causes it to leave (in which
case it will eventually hit the other wall). We shall see that swimmer-wall hydrodynamic
interactions have a quantitative effect on the layering, but not a qualitative one. Recent
experiments with bacteria also show a very high concentration at the
walls~\cite{LaugaSurfaceAttractPRL08}. Because the formation of layers is a generic phenomenon, the
present theory focuses on the dynamics of swimmers in the layers.

In the absence of other swimmers, a swimmer will continue in a straight line within the layer.
Collisions or hydrodynamic interactions with other swimmers cause the direction of swimming to
change, leading to diffusive motion at long time scales. From a scaling perspective, consider a
random walk in which the particle swims at constant speed $v_\mathrm{s}$ along a trajectory for a
mean duration $\tau_\mathrm{s}$ before changing directions. The motion at long times is diffusive
with a diffusivity that scales as $D_\mathrm{s} \sim v_\mathrm{s}^2 \tau_\mathrm{s} \sim
v_\mathrm{s} l_\mathrm{s}$, where $l_\mathrm{s} = v_\mathrm{s} \tau_\mathrm{s}$ is the mean free
path. In a dilute system, the velocity of the swimmer is essentially the isolated swimmer value,
i.e. $v_\mathrm{s} \sim v_{\mathrm{is}}$. (Changes in swimming speed due to confinement do not affect the scaling predictions.) Deviations from this at larger concentrations will be discussed later. The mean
free path scales as $l_\mathrm{s} \sim \ell^2 ( \sigma \psi_\mathrm{e} )^{-1}$, where $\sigma$ is
a two-dimensional cross section (with units of length) for the redirections. Therefore, the
swimmer diffusivity, $D_\mathrm{s}$, will scale as
\begin{equation}
D_\mathrm{s} \sim (\sigma \psi_\mathrm{e})^{-1}, \label{eq:scaling1}
\end{equation}
in dilute systems. Analysis for an unconfined domain~\cite{underhill-spp} yields a similar scaling
but with a three-dimensional cross section $\sigma_{\mathrm{3D}}$ and the area fraction replaced
by a volume fraction, so $D_{\mathrm{s,3D}} \sim (\sigma_{\mathrm{3D}} \phi_\mathrm{e})^{-1}$.

We now turn to the fluid flow generated by the motion of the swimmers, considering the behavior of
passive, non-Brownian tracers that follow the local fluid velocity. The tracers undergo
ballistic motion at short lag times and diffusive motion at large lag times. The velocity of a
tracer, ${\bf v}_\mathrm{t}$, is the fluid velocity at the location of the tracer ${\bf
x}_\mathrm{t}$, and is the sum of the disturbance velocities due to each swimmer:
\begin{equation}
{\bf v}_\mathrm{t} = {\bf u}({\bf x}_\mathrm{t}) = \sum_{i=\nu}^N {\bf u}^\mathrm{d}_\nu ,
\label{eq:scaling2}
\end{equation}
where ${\bf u}^\mathrm{d}_\nu$ is the disturbance at the position of the tracer due to swimmer
$\nu$. We can calculate the mean-squared velocity of the tracers by squaring this sum and
performing an ensemble average over all possible configurations of the swimmers while the position
of the tracer is held fixed at $\mathbf{x}_{\mathrm{t}}$:
\begin{equation}
\llangle {\bf v}_\mathrm{t} \cdot{\bf v}_\mathrm{t} \rrangle_{ \mathbf{x}_{\mathrm{t}}} = \llangle
v_\mathrm{t}^2 \rrangle_{ \mathbf{x}_{\mathrm{t}}} = \sum_{\nu=1}^N \sum_{\mu=1}^N \llangle {\bf
u}^\mathrm{d}_\nu \cdot {\bf u}^\mathrm{d}_\mu \rrangle_{ \mathbf{x}_{\mathrm{t}}} ,
\end{equation}
where the subscript $\mathbf{x}_{\mathrm{t}}$ denotes that the tracer is held fixed at
$\mathbf{x}_{\mathrm{t}}$ during the ensemble average over swimmers. The ensemble average without
such a subscript means an average over all tracer positions has also been performed. In the dilute
limit, we assume the swimmers to be distributed independently. This means that each swimmer is
independently sampling a probability distribution for location within the channel, though that
probability distribution need not be uniform. For example, we consider here that distribution to
be peaked in a layer near each wall. The independence of the swimmers leads to
\begin{equation}
\llangle v_\mathrm{t}^2 \rrangle_{\mathbf{x}_{\mathrm{t}}} \sim N \llangle \left( u^\mathrm{d}_\nu
\right)^2 \rrangle_{\mathbf{x}_{\mathrm{t}}} \sim \phi_\mathrm{e} \sim \psi_\mathrm{e}  .
\label{eq:scaling3}
\end{equation}
The number of swimmers $N$ is proportional to both the volume fraction and the area fraction with
a different proportionality factor that depends on the channel height.

A similar scaling analysis can be used for the tracer diffusivity, which can be written using the
Green-Kubo relation~\cite{reichl,mcquarrie}
\begin{equation}
D_\mathrm{t} = \frac{1}{3} \int_0^\infty \llangle {\bf v}_\mathrm{t}(0)\cdot {\bf v}_\mathrm{t}(t)
\rrangle \rmd t . \label{eq:gree-kubo}
\end{equation}
Again, we can replace the tracer velocity by a sum over swimmer disturbances. Assuming independent
swimmers in the dilute limit gives
\begin{equation}
D_\mathrm{t} \sim \frac{N}{3} \int_0^\infty \llangle {\bf u}^\mathrm{d}_\nu(0) \cdot {\bf
u}^\mathrm{d}_\nu(t) \rrangle \rmd t \sim \phi_\mathrm{e} \sim \psi_\mathrm{e} .
\end{equation}
Combining the velocity and diffusivity scalings with the definition of a correlation time,
$\tau_\mathrm{t} = D_\mathrm{t} /\langle v_\mathrm{t}^2 \rangle$ gives a prediction that
$\tau_\mathrm{t}$ is independent of volume fraction in the dilute limit.

We can also use this behavior of the tracers to understand the correction to the swimmer behavior
at finite concentration. The tracers represent a characteristic fluid element in the system. In
addition to being self-propelled, each swimmer is advected by the local fluid disturbance
(generated by the other swimmers). If the fluid velocity does not change rapidly over the size of
a swimmer, then we can write
\begin{equation}
\mathbf{v}_\mathrm{s} \approx \mathbf{v}_{\mathrm{is}} + \mathbf{v}_\mathrm{t} (
\mathbf{x}_\mathrm{s} ) .\label{eq:vsum}
\end{equation}
Squaring this expression and neglecting the cross terms in the dilute limit gives
\begin{equation}
\llangle v_\mathrm{s}^2 \rrangle - v^2_{\mathrm{is}} \sim \llangle v_\mathrm{t}^2 \rrangle .
\label{eq:scaling4}
\end{equation}
Note that there exists a proportionality factor because the average over the fluid in the region
occupied by the swimmers is different than the average over the fluid in the whole domain. The
proportionality is replaced by an equality for an unconfined system, for which these domains are
the same~\cite{underhill-spp}. Combining this result with the scaling of the tracer velocity
$\langle v_\mathrm{t}^2 \rangle \sim \phi_\mathrm{e} \sim \psi_\mathrm{e}$ gives that:
\begin{equation}
\llangle v_\mathrm{s}^2 \rrangle - v^2_{\mathrm{is}} \sim \phi_\mathrm{e} \sim \psi_\mathrm{e} .
\end{equation}
and similarly that
\begin{equation}
\llangle v_\mathrm{s}^2 \rrangle^{1/2} - v_{\mathrm{is}} \sim  \phi_\mathrm{e} \sim
\psi_\mathrm{e} . \label{eq:scaling5}
\end{equation}

\subsection{Velocity fields and spatial correlations}
An important measure of correlations in a suspension of swimming microorganisms is the spatial
autocorrelation function of the fluid velocity. We will report results for correlations on planes
of constant $x_3$. The separation of different fluid elements in this plane is denoted by ${\bf
x}_{\|}$. Therefore, the fluid correlation is $C_{\mathrm{f}}({\bf x}_{\|},x_3)=\llangle {\bf
u}({\bf s}_{\|},x_3)\cdot {\bf u}({\bf s}_{\|} + {\bf x}_{\|},x_3) \rrangle$, where the angle
bracket represents an ensemble average over swimmers and also an average over the tracer position
 ${\bf s}_{\|}$ in the plane $x_3=\textrm{const.}$ 

A number of experimental studies \cite{GoldsteinKesslerSPP04,KesslerZoomingNematicExpFluids2007} have reported the spatial correlation function for the swimmer velocities: 
$C_{\mathrm{s}}({\bf x}_{\|},x_3)=\llangle \mathbf{v}_\mathrm{s}({\bf s}_{\|},x_3)\cdot \mathbf{v}_\mathrm{s}({\bf s}_{\|} + {\bf x}_{\|},x_3) \rrangle$. 
Assuming the validity of the approximation given by Eq.~\ref{eq:vsum}, we can write that
\begin{equation}
\eqalign{C_{\mathrm{s}}({\bf x}_{\|},x_3) &=\llangle \mathbf{v}_\mathrm{is}({\bf s}_{\|},x_3)\cdot \mathbf{v}_\mathrm{is}({\bf s}_{\|} + {\bf x}_{\|},x_3) \rrangle \cr
 & + 2\llangle \mathbf{v}_\mathrm{is}({\bf s}_{\|},x_3)\cdot {\bf u}({\bf s}_{\|} + {\bf x}_{\|},x_3) \rrangle \cr
 & + C_{\mathrm{f}}({\bf x}_{\|},x_3).}
\end{equation}
 In the case of independent swimmers whose orientations are uncorrelated with the fluid velocity, this expression reduces to 
\begin{equation}
C_{\mathrm{s}}({\bf x}_{\|},x_3) =v_{\mathrm{is}}^2\chi_0({\bf x}_{\|})+ C_{\mathrm{f}}({\bf x}_{\|},x_3),
\end{equation}
where $\chi_0({\bf x}_{\|})=1$ if ${\bf x}_{\|}=0$ and zero otherwise. This expression indicates the close relationship between $C_{\mathrm{s}}$ and $C_{\mathrm{f}}$. In our simulations, $C_{\mathrm{f}}$ is substantially less susceptible to statistical noise than $C_{\mathrm{s}}$ so that is the quantity we report.

The correlation function $C_{\mathrm{s}}$ has been used in experiments and simulations
to quantify the ``swirls'' seen in the
fluid~\cite{GoldsteinKesslerSPP04,KesslerZoomingNematicExpFluids2007}. In particular, the
existence of negative correlations has been used as evidence for collective behavior and used to
quantify the size of the swirls. However, we will see here that negative correlations can be
present even in the absence of collective behavior. The presence of the walls changes the
disturbance that an organism produces in the fluid, and thus the correlations in the fluid.

The spatial fluid correlations are determined by the fluid disturbance caused by a single swimmer
as well as correlations between the swimmers. We consider here the correlations present in the
absence of correlations between the swimmers, that is for independent swimmers. The correlation
for independent swimmers is determined solely by the disturbance caused by a single swimmer and
the concentration of swimmers. Recall from figure~1 the disturbance that a single swimmer produces
in the fluid in the absence of walls, a pusher expels fluid out from the front and back, while
sucking in fluid from the sides. However, the streamlines do not form closed loops, or swirls.
Contrast this with figure~2, which shows the streamlines of the disturbance produced by a single
swimmer in the presence of walls. The walls induce swirls in the flow. Two new length scales, the
separation of the walls and the separation of a swimmer from the walls, affect the fluid
structures.

\begin{figure}
\begin{center}
\includegraphics[width=0.47\textwidth]{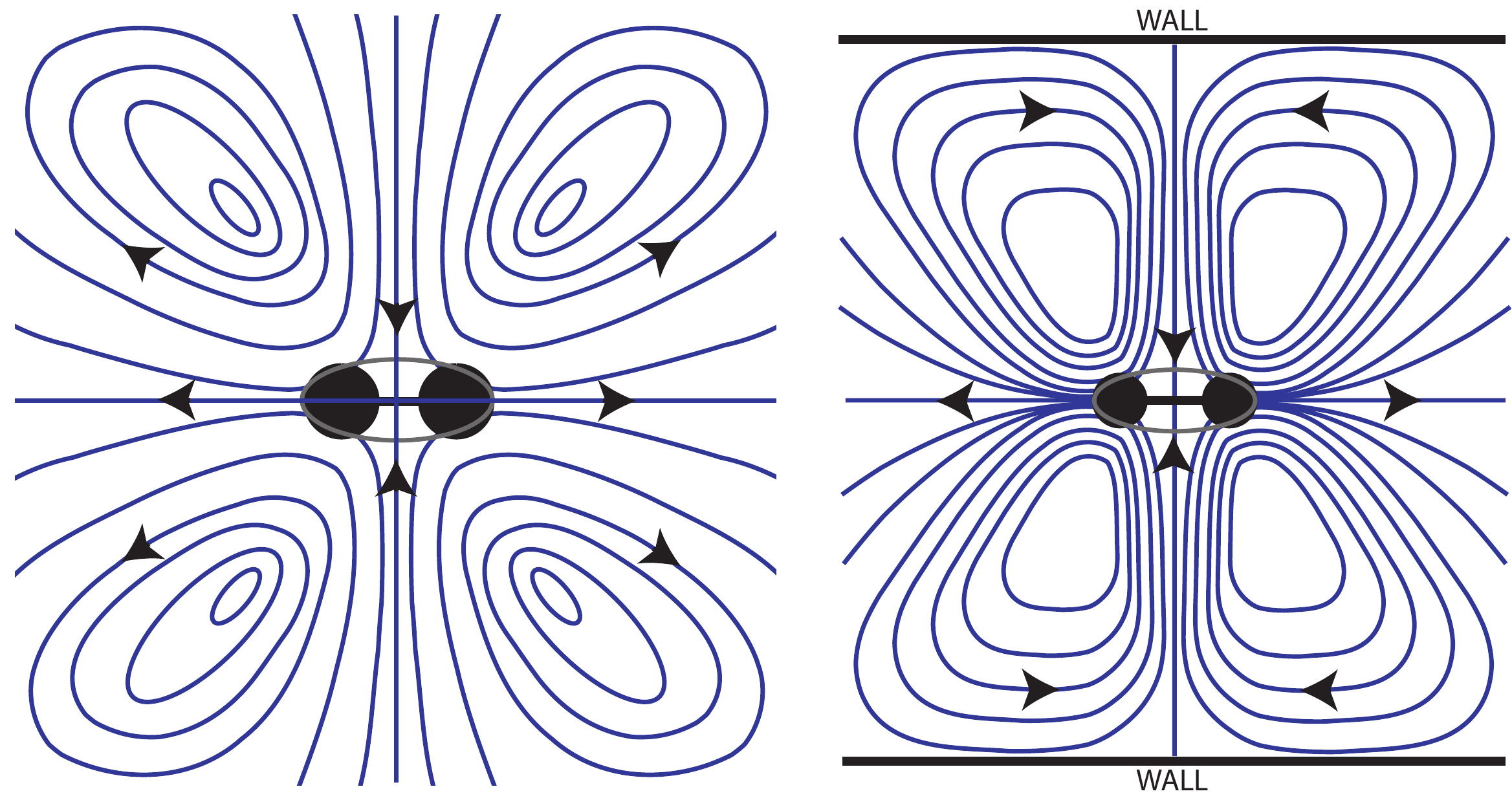}\\
\includegraphics[width=0.47\textwidth]{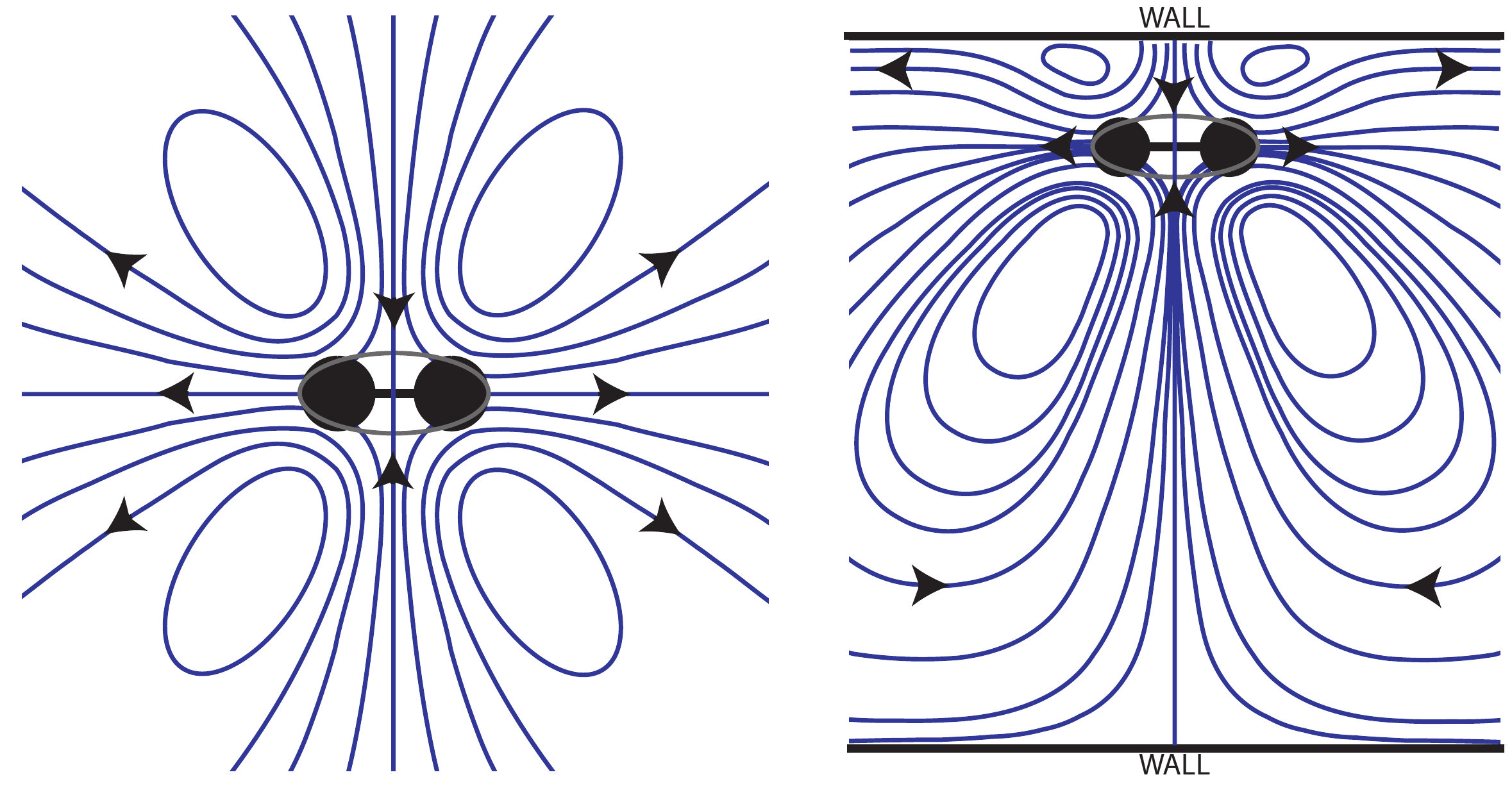}
\caption{Streamlines of the fluid disturbance due to a force dipole representing a pusher: at the
middle of the channel (top row) and close to the wall (bottom row). The fluid is shown in the
$x_1$-$x_2$ plane (left column) and the $x_1$-$x_3$ plane (right column). }
\end{center}
\label{fig:swimmer-wall}
\end{figure}

\begin{figure}
\begin{center}
\includegraphics[width=0.47\textwidth]{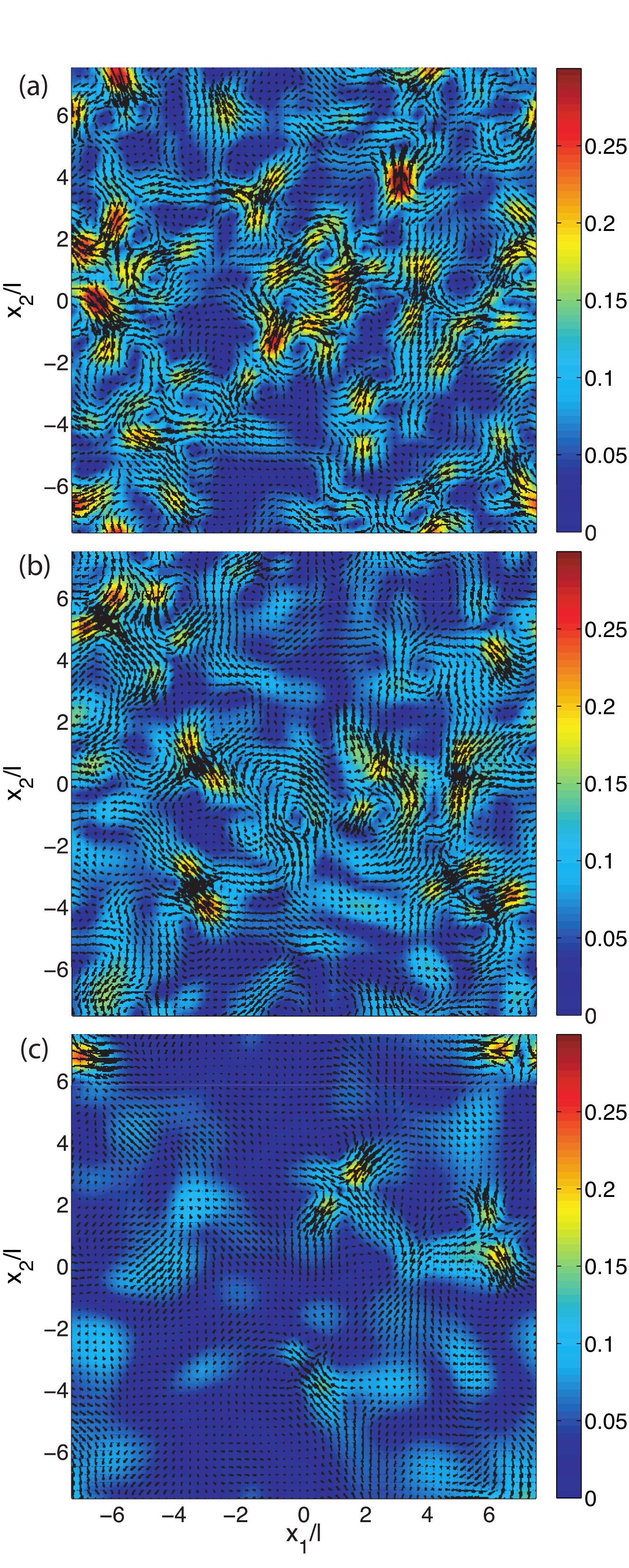}
\caption{Snapshots of the velocity field for independent swimmers in layers at
$\phi_{\mathrm{e}}=0.1$ ($\psi_{\mathrm{e}}=0.375$) with $L=15 \ell$ and $2H=5 \ell$. (a) $x_3=0.8
H \ell$ (b) $x_3=0.5 H \ell$ (c) $x_3=0$  } \label{fig:u-fields6}
\end{center}
\end{figure}

The difference between these flow fields leads to different fluid correlations. As we show below,
the swirl in the fluid induced by the wall means that even independent swimmers, that have no
collective behavior, will produce fluid correlations that are negative. To illustrate this point,
it is useful to visualize snapshots of a typical velocity field generated by these independent
swimmers, as shown in figure~\ref{fig:u-fields6}. It is important to note that these snapshots are
simply a sum of disturbances due to independent swimmers, each producing a disturbance like that
in figure~2. The snapshots look remarkably similar to the results of simulations and experiments
for which the swirls were considered evidence of collective behavior.

\begin{figure}
\begin{center}
\includegraphics[width=0.47\textwidth]{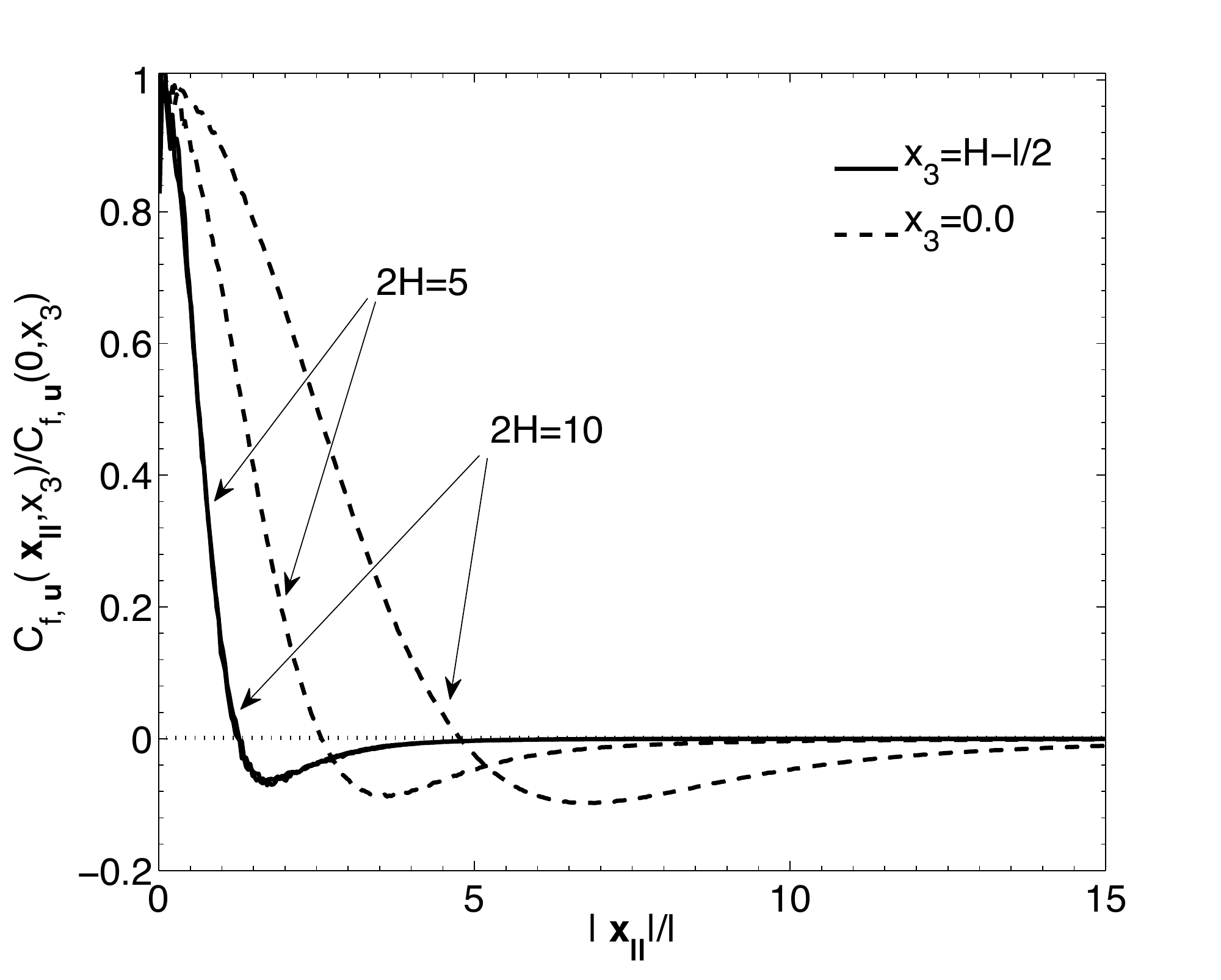}
\caption{Fluid velocity autocorrelation function for independent swimmers at $2H=5 \ell$ and
$2H=10 \ell$. The autocorrelation is normalized by the corresponding value at $|{\bf x}_{||} |=0$.
When $x_3=H-\ell/2$ the curves for $2H=5 \ell$ and $2H=10 \ell$ are indistinguishable. }
\label{fig:auto-u-size-ind}
\end{center}
\end{figure}

The autocorrelation function within a fluid for which there are such swirls has a region of
negative correlation, which is shown in figure~\ref{fig:auto-u-size-ind}. The correlation function
has been calculated in two planes, the plane in which the swimmers form layers as well as the
center plane of the slit. By changing the amount of confinement, we see that the length scale at
which the correlation becomes negative in the center of the slit is the separation of the walls.
However, the correlation near the wall, in the plane of the swimmers, is independent of the
confinement. Instead, the length scale depends on the size of a swimmer and the separation of the
swimmers from the wall. This is true for the cases shown because the second wall is far enough
away that its effect on the fluid correlation is small.

By examining theoretically the fluid correlations generated by uncorrelated swimmers, we have
shown that collective behavior is not necessary to generate swirls and negative correlations in
the fluid. This illustrates the importance of comparing properties calculated from a suspension of
swimming organisms to that of independent organisms to gauge the importance of collective behavior
in producing the observed response.

\section{Computational Results}\label{sec:results}
The theory just described makes distinct predictions regarding swimmer and tracer diffusivities,
velocities, and fluid phase correlations. In this section we directly compute these quantities and
others for suspensions over a range of concentrations and degrees of confinement. For the
remainder of the article, all lengths are non-dimensionalized by the swimmer size $\ell$ and all
times are non-dimensionalized by $\ell/v_{\mathrm{is}}$.

\begin{figure}
\begin{center}
\includegraphics[width=0.47\textwidth]{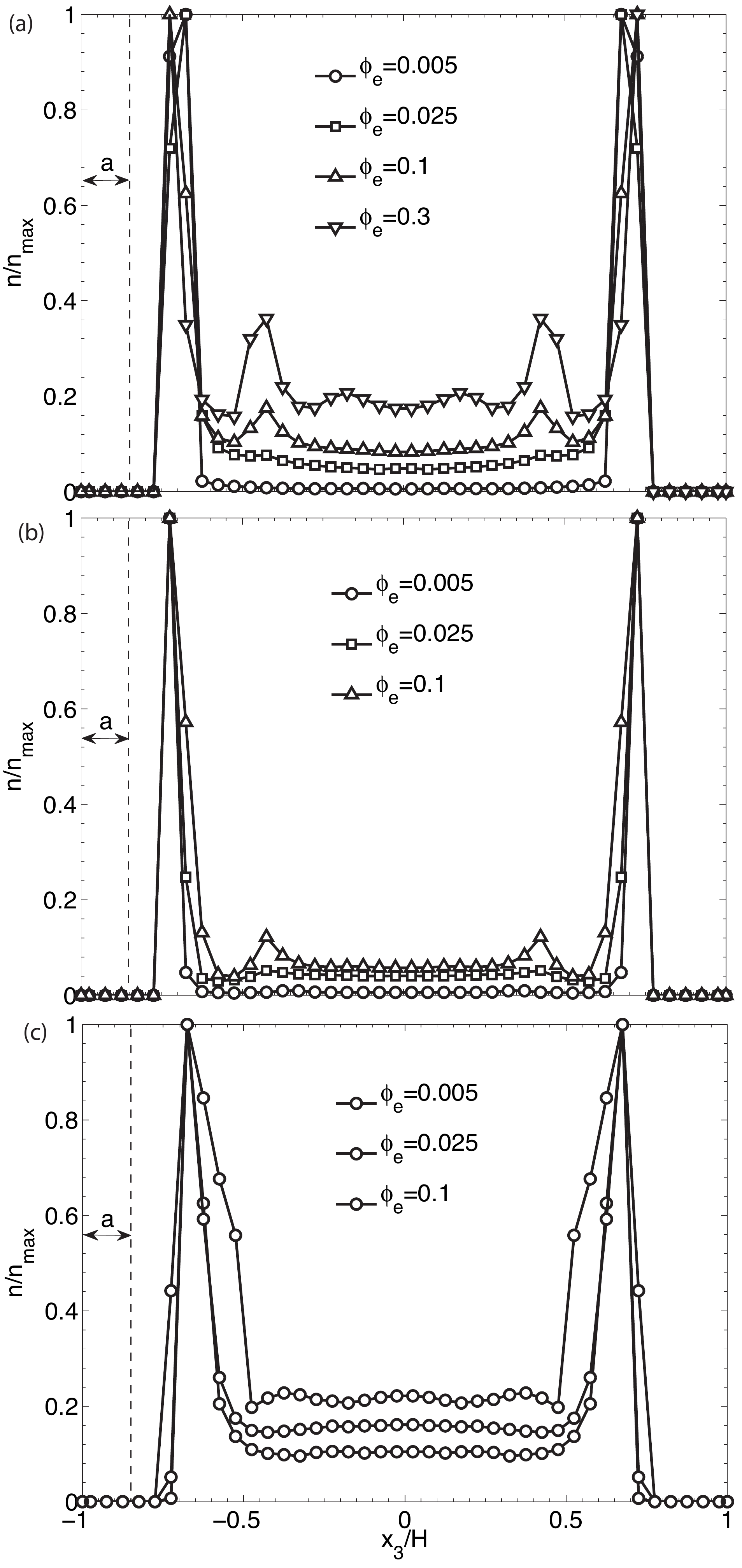}
\caption{Concentration profile as a function of wall-normal position $x_3$ for different effective
volume fractions with $L = 15$ and $2H = 5$. (a) pushers (b) pushers without hydrodynamic
interactions (c) pullers  } \label{fig:rho-h5}
\end{center}
\end{figure}

Let us start by fixing the confinement to five swimmer sizes, $2H=5$, and the periodicity at three
times the confinement, $L=3\times2H$. We will show later that this periodicity is large enough for
the results to be independent of $L$ because the walls prevent formation of flow structures larger
than the confinement. In the dilute theory described previously, it was assumed that the swimmers
formed a layer at each wall. The formation of layers close to the confining wall in the
simulations is illustrated in figure~\ref{fig:rho-h5}, which shows the swimmer concentration
profiles as functions of $x_3$ at different effective volume fractions. Results for both pushers
and pullers are shown, as well as for swimmers where hydrodynamic effects are completely
neglected. For very dilute systems, the results in all cases are fairly similar -- layers form
near each wall and there is a nearly zero concentration at small distances from the walls due to
steric exclusion of beads from the walls. However, the structure within the layer does depend on
whether hydrodynamic interactions are included. With hydrodynamic interactions, no long-range
orientational order is observed in the layer. Without hydrodynamic interactions, a two-dimensional
nematic state is observed, similar to the three-dimensional nematic state observed in an
unconfined system without hydrodynamic interaction~\cite{underhill-spp}. Once the concentration is
high enough that the layer close to the wall is saturated, the swimmers form additional layers
indicated by the secondary peaks in the profiles at high concentrations.

\begin{figure}
\begin{center}
\includegraphics[width=0.47\textwidth]{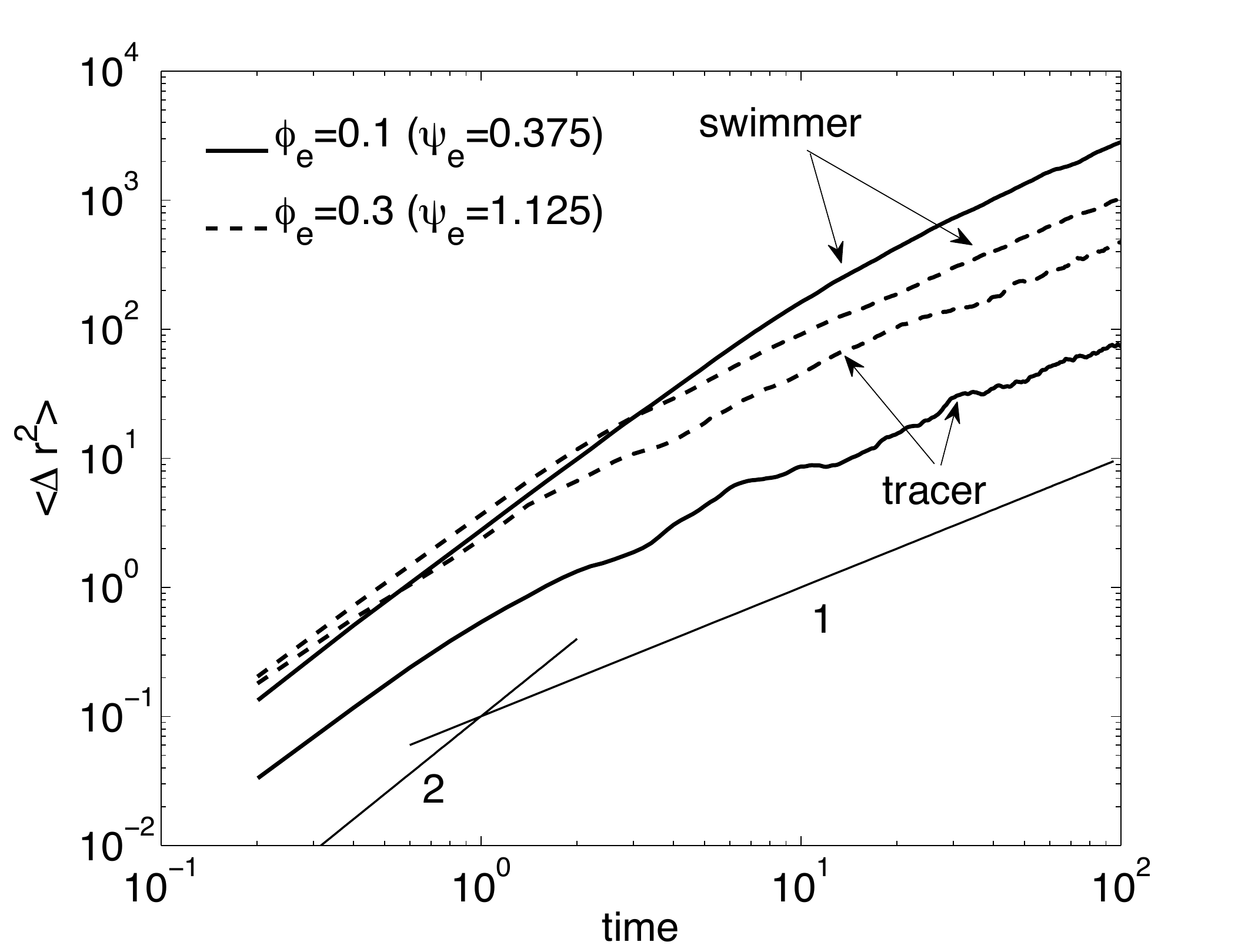}
\caption{Mean-squared-displacement vs time for swimmers and tracers at various concentrations with
$L = 15$ and $2H = 5$. } \label{fig:msd-h5}
\end{center}
\end{figure}

\begin{figure}
\begin{center}
\includegraphics[width=0.47\textwidth]{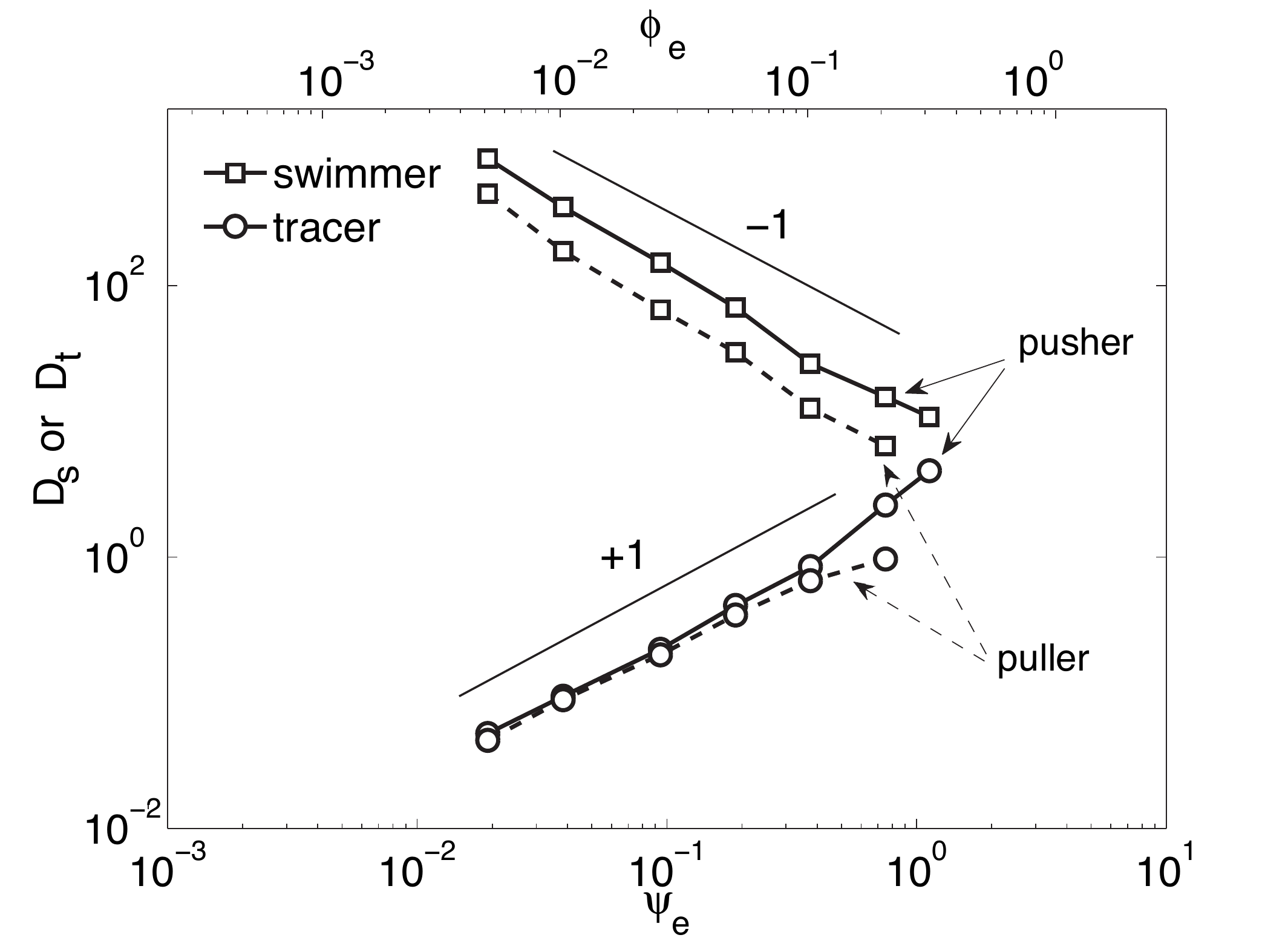}
\caption{Diffusion coefficients as a function of the effective volume fraction (top axis) and
effective area fraction (bottom axis) for swimmers and tracers with $L = 15$ and $2H = 5$. }
\label{fig:d-h5}
\end{center}
\end{figure}

Figure~\ref{fig:msd-h5} shows the mean-squared-displacement (MSD) in the periodic ($x_1 , x_2$)
plane as a function of time, for swimmers and fluid tracers, at two concentrations:
$\phi_{\mathrm{e}}=0.1$ ($\psi_{\mathrm{e}}=0.375$) and $\phi_{\mathrm{e}}=0.3$
($\psi_{\mathrm{e}}=1.125$). At short times, the MSD is ballistic, reflecting the straight-line
motions of an isolated swimmer. The duration of this ballistic regime decreases as the
concentration increases. At longer times the behavior becomes diffusive. Since Brownian motion is
absent, the origin of this diffusive regime is the interactions between the swimmers.
Figure~\ref{fig:d-h5} shows the long-time diffusion coefficients $D_\mathrm{s}$ and $D_\mathrm{t}$
as a function of the effective volume fraction and area fraction. At low concentrations the
swimmers have a high effective diffusion coefficient reflecting the weak hydrodynamic interactions
between them.  The flow is only disturbed by a small number of swimmers so tracers diffuse very
slowly. As the concentration is increased the diffusivity of the swimmers decreases, as the
natural ballistic trajectories are increasingly perturbed by hydrodynamic interactions and
collisions with other swimmers. Correspondingly, the naturally motionless tracers feel the motion
of increasingly more swimmers and their diffusivity increases.

According to the results in figure~\ref{fig:d-h5}, the diffusion coefficient for both swimmers and
tracers follow the dilute theory scalings ( $ D_\mathrm{s} \sim \psi_{\mathrm{e}}^{-1} $ , $
D_\mathrm{t} \sim \psi_{\mathrm{e}} $ ) developed in section~\ref{sec:theory} even at moderate
concentrations. Note the difference in diffusivity and density profiles compared to our previous
work~\cite{hernandez-spp}. For the volume fractions shown in the present work, we do not see the
same dramatic increase in diffusivity and shift in the density profiles to the center of the
channel. These differences seem to be due to the use of regularized forces in the present work,
versus point forces in the former, and an excluded volume potential as shown in figure~1. With
this excluded volume potential and swimmer aspect ratio, obtaining much larger volume fractions is
not possible because there is a largest volume fraction corresponding to the close-packed state.

\begin{figure}
\begin{center}
\includegraphics[width=0.47\textwidth]{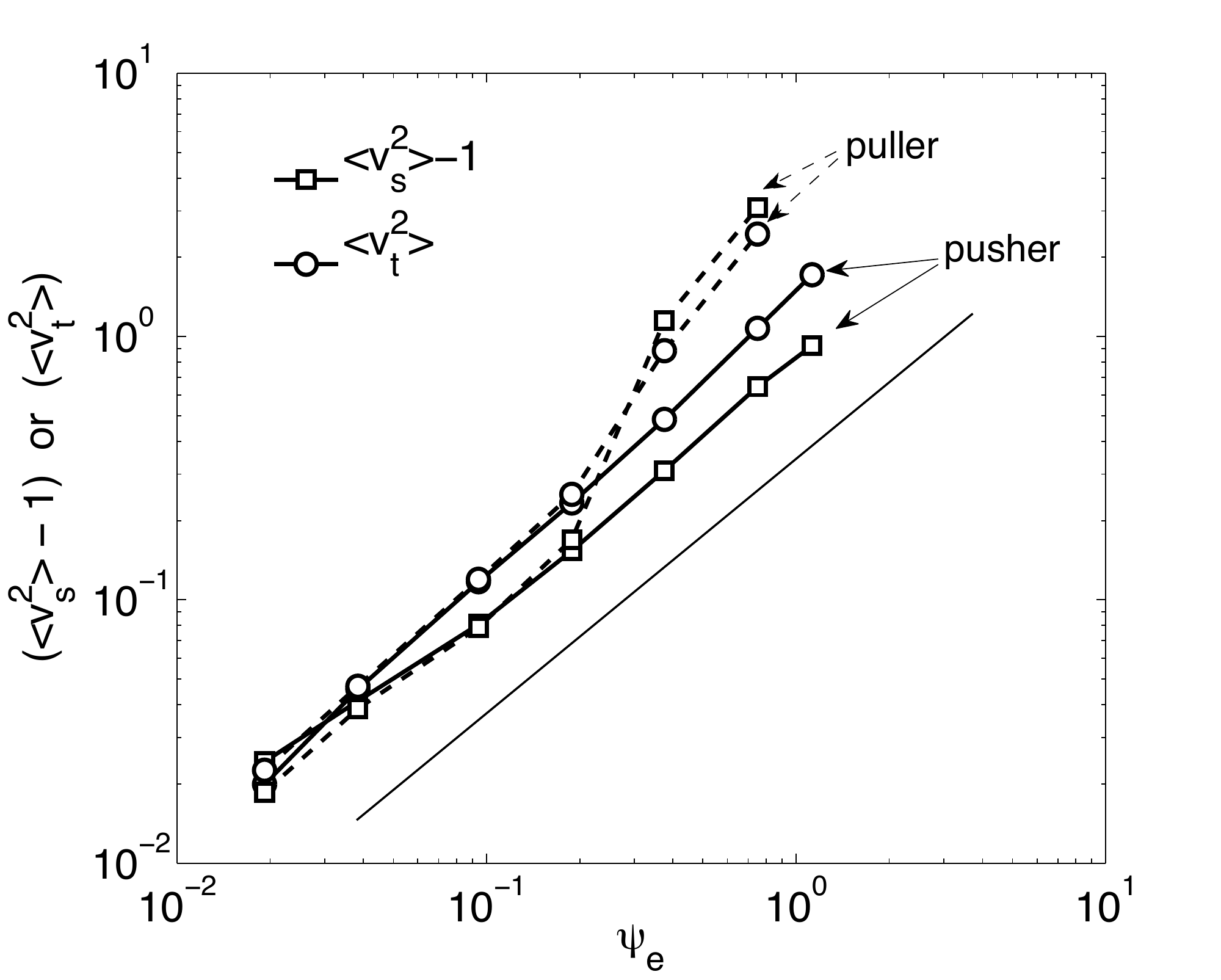}
\caption{Mean-squared velocity for swimmers, $\llangle v^2_s \rrangle - 1$, and tracers, $\llangle
v^2_t \rrangle$ as a function of the effective area fraction for $L = 15$ and $2H = 5$. }
\label{fig:u-h5}
\end{center}
\end{figure}

From section~\ref{sec:theory}, we saw that the diffusivities in figure~\ref{fig:d-h5} are related
to the typical velocities of the swimmers and tracers. Figure~\ref{fig:u-h5} shows $\langle
v^2_\mathrm{t} \rangle$ and $\llangle v^2_\mathrm{s} \rrangle - 1$ as functions of
$\psi_{\mathrm{e}}$. Both display an approximately linear increase with concentration, as
predicted by the simple theory. We believe the deviation at small concentration is because of
statistical errors due to the small number of swimmers in the domain at small concentrations. As
with the diffusivities, the dilute scaling seems to hold even at concentrations well above
$\psi_{\mathrm{e}} = 0.1$ for pushers. However, the puller simulations deviate from the dilute
scaling at $\psi_{\mathrm{e}} \gtrsim 0.2$. The cause for this difference between pushers and
pullers is unknown.

\begin{figure}
\begin{center}
\includegraphics[width=0.47\textwidth]{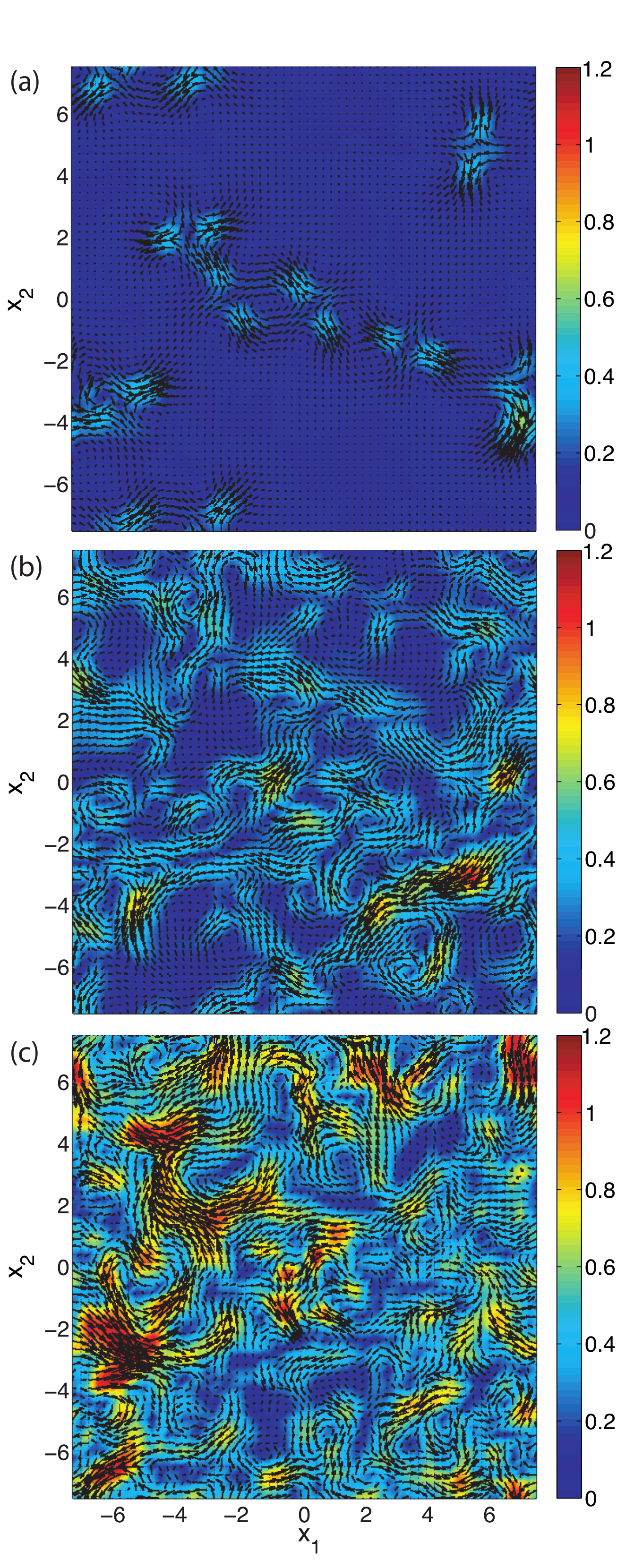}
\caption{Snapshots of the velocity field at $x_3=0.8H$  ($x_3=H-1/2$) with $L=15$ and $2H=5$. (a)
$\phi_{\mathrm{e}}=0.01$ ($\psi_{\mathrm{e}}=0.0383$) (b) $\phi_{\mathrm{e}}=0.1$
($\psi_{\mathrm{e}}=0.375$) (c) $\phi_{\mathrm{e}}=0.3$ ($\psi_{\mathrm{e}}=1.125$)  }
\label{fig:u-fields1-wall}
\end{center}
\end{figure}

\begin{figure}
\begin{center}
\includegraphics[width=0.47\textwidth]{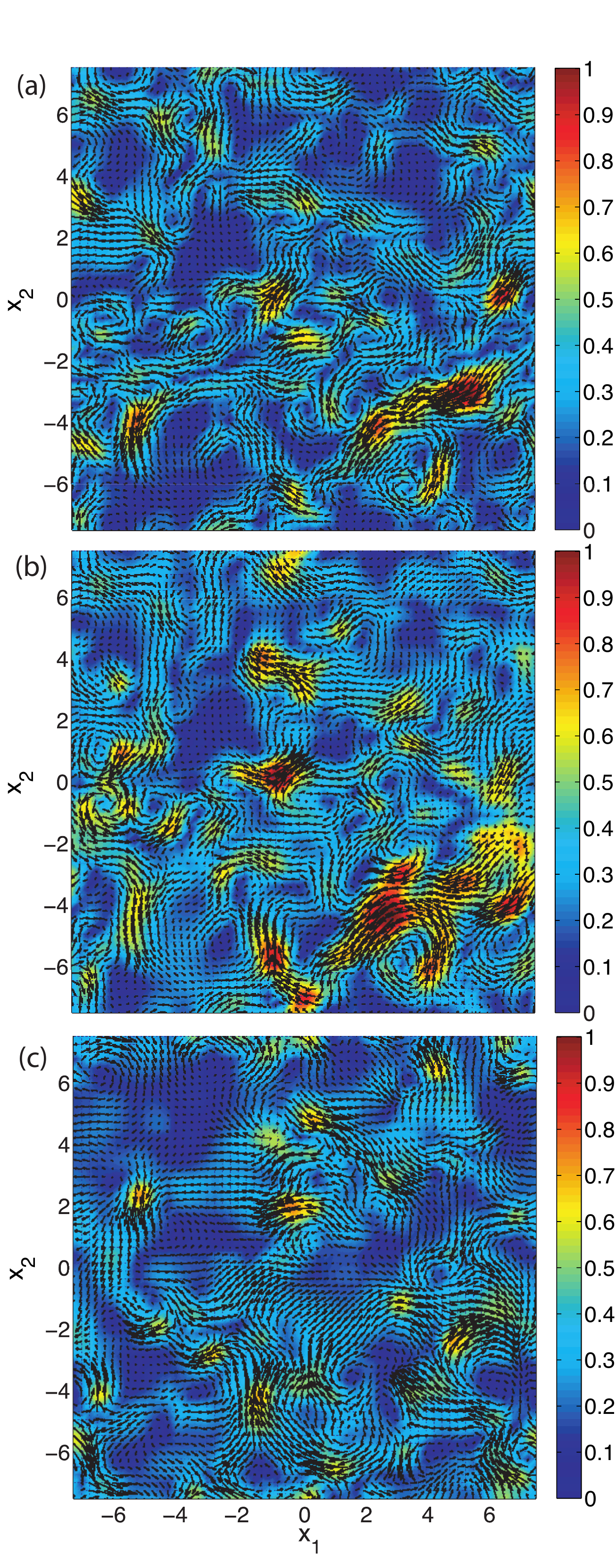}
\caption{Snapshots of the velocity field for $\phi_{\mathrm{e}}=0.1$ ($\psi_{\mathrm{e}}=0.375$)
with $L=15$ and $2H=5$. (a) $x_3=0.8H$ ($x_3=H-1/2$) (b) $x_3=0.5H$ (c) $x_3=0$ }
\label{fig:u-fields2}
\end{center}
\end{figure}

The final property we examine for this initial confinement is the fluid flow generated by the
swimmers. Figure~\ref{fig:u-fields1-wall} shows snapshots of the fluid velocity field after more
than 1000 dimensionless times in a plane near the wall of the channel ($x_3=0.8 H$) for different
concentrations $\phi_{\mathrm{e}}=0.01, 0.1$ and $0.3$ ($\psi_{\mathrm{e}}=0.0384, 0.375$ and
$1.125$). These velocity fields are similar to those observed in some
experiments~\cite{GoldsteinKesslerSPP04}. Snapshots suggest that the size of the flow structures
is related to the wall separation $2H$ (as we will address later). We also examined the snapshots
for three different locations in the channel, $x_3 = 0, 0.5H$, and $0.8H$, which are shown in
figure~\ref{fig:u-fields2} for $\phi_{\mathrm{e}}=0.1$ ($\psi_{\mathrm{e}}=0.375$).

\begin{figure}
\begin{center}
\includegraphics[width=0.47\textwidth]{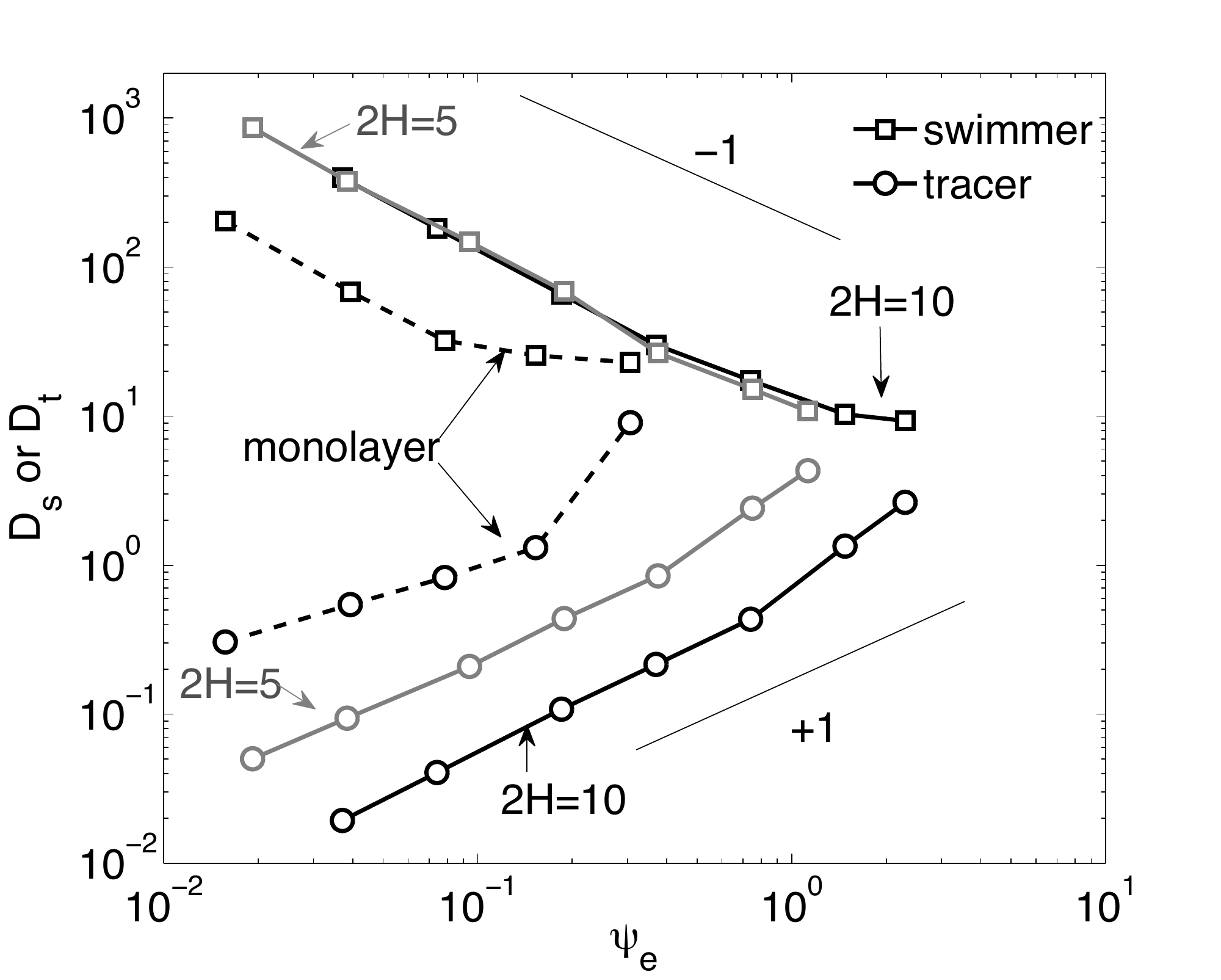}
\caption{Diffusion coefficients vs effective area fraction for swimmers and tracers for different
confinements. For $2H=5$ and $2H=10$ we used $L=3\times 2H$, while for the monolayer $2H=1$ we
used $L=20\times 2H$. } \label{fig:d-h10}
\end{center}
\end{figure}

\begin{figure}
\begin{center}
\includegraphics[width=0.47\textwidth]{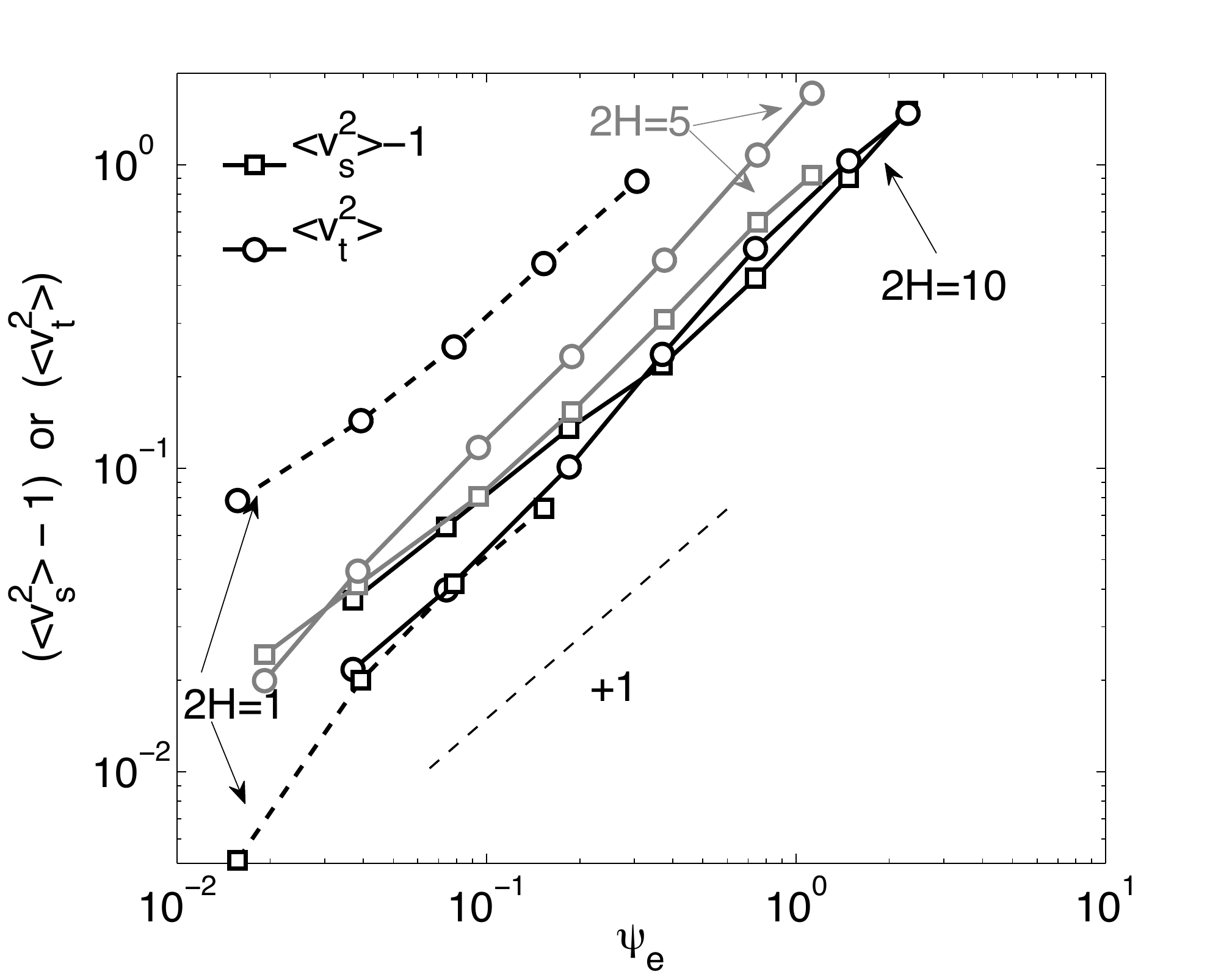}
\caption{Mean-squared velocity for swimmers, $\llangle v^2_\mathrm{s} \rrangle - 1$, and tracers,
$\llangle v^2_\mathrm{t} \rrangle$ vs effective area fraction. For $2H=5$ and $2H=10$ we used
$L=3\times 2H$, while for the monolayer $2H=1$ we used $L=20\times 2H$. } \label{fig:u-h10}
\end{center}
\end{figure}

We now turn to the effect of the degree of confinement on the results. Simulations with a wider
gap ($2H=10$ and $L=3\times2H$) will be examined as well as a highly confined case of $2H=1$,
$L=20\times2H$, and with the swimmers restricted to lie in the midplane of the channel -- a
monolayer. Figures~\ref{fig:d-h10} and \ref{fig:u-h10} show the long-time diffusion coefficients
$D_\mathrm{t}$ and $D_\mathrm{s}$, the mean-squared swimmer velocity deviation $\llangle
v^2_\mathrm{s} \rrangle - 1$, and the mean-squared fluid velocity $\llangle v^2_\mathrm{t}
\rrangle$ for all three confinements. The first observation is that all confinements follow the
dilute theory scalings at low concentrations. In particular, we see that the swimmer diffusivities
for $2H=5$ and $2H=10$ confinements collapse reasonably well when the concentration is represented
as an effective area fraction. This supports the simple theory that assumes the swimmers form a
layer next to each wall in the dilute limit. The swimmer diffusivities in the monolayer have the
same scaling, though a different prefactor. This difference is most likely due to either a
difference in the hydrodynamic interactions between swimmers in the monolayer because of the
nearby walls or the inability of the swimmers to escape from the layer.

The tracer diffusivities, while following the simple scaling with area fraction, do not collapse
with $\psi_{\mathrm{e}}$ with changing confinement. Because the tracers are not restricted to
planes near the walls, at weak confinement (large $H$) the tracers can sample the lower velocities
in the center of the channel, away from the swimmer layers near the walls. This difference results
in a lower velocity of tracers with weaker confinement, as seen in figure~\ref{fig:u-h10}.

\begin{figure}
\begin{center}
\includegraphics[width=0.47\textwidth]{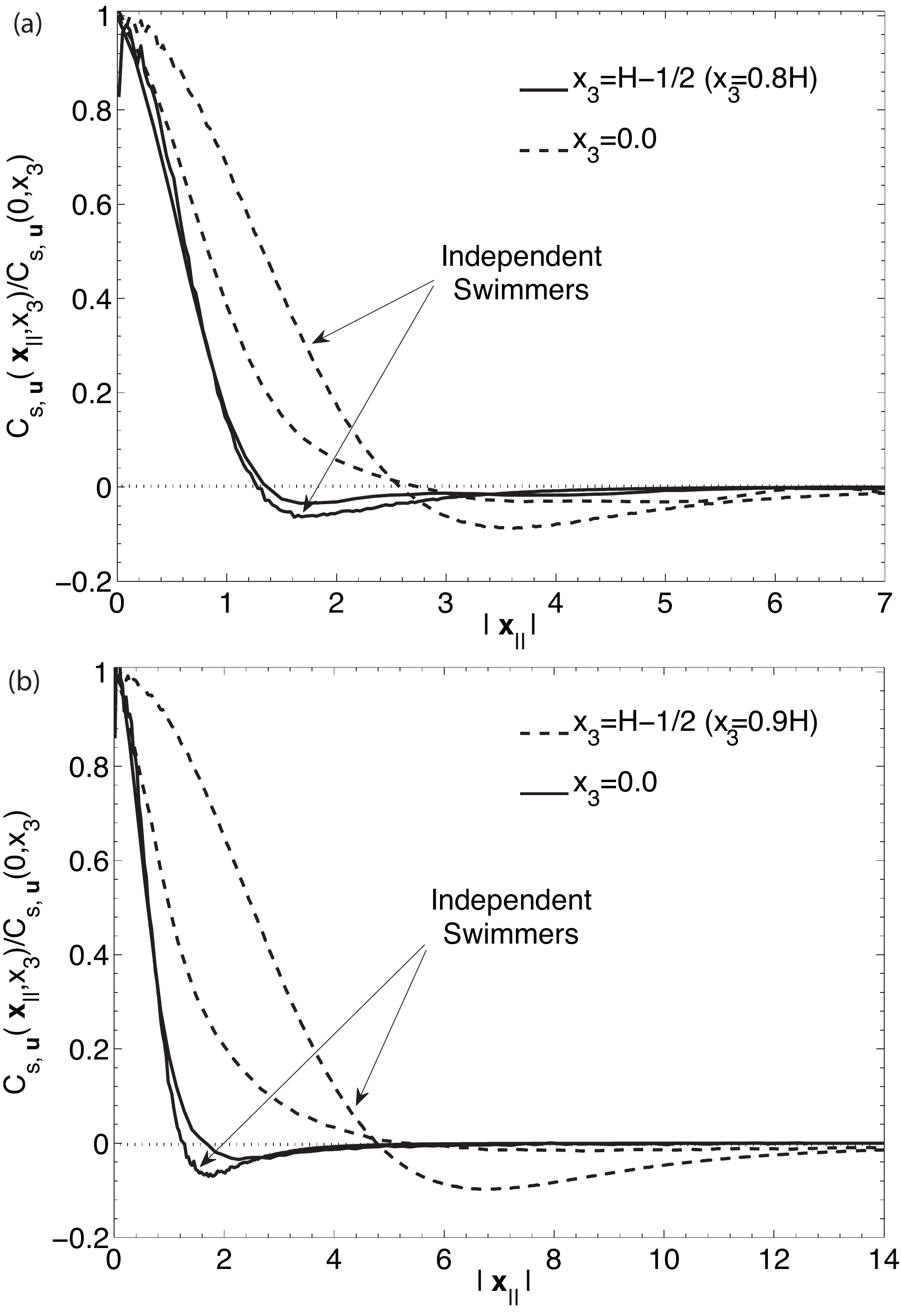}
\caption{Fluid velocity autocorrelation function at $\phi_{\mathrm{e}}=0.05$ (a) $2H=5$
($\psi_{\mathrm{e}}=0.19$) (b) $2H=10$ ($\psi_{\mathrm{e}}=0.37$) } \label{fig:auto-u}
\end{center}
\end{figure}

From the earlier theoretical analysis, we found that uncorrelated swimmers produce swirls in the
fluid, resulting in negative fluid correlations. The size of these negative regions were related
to the separation of the walls, the size of a swimmer, and the separation of a swimmer from a
wall. We found in the simulations swirls in the fluid as shown in figures~\ref{fig:u-fields1-wall}
and \ref{fig:u-fields2}. To make a quantitative comparison between the simulations and the theory
of uncorrelated swimmers, we compare the fluid autocorrelation functions in planes parallel to the
slit. Figure~\ref{fig:auto-u} shows the velocity autocorrelation function at two $x_3$-planes,
$x_3=0.0$ and $0.8H$, for $2H=5$ and $2H=10$, both with $L=3\times 2H$ and $\phi_{\mathrm{e}}
=0.05$. These results indicate, as foreshadowed by the velocity fields, that the scale of the
velocity autocorrelation is set by the distance between the walls. The fluid velocities are
correlated at short length scales and, before they become decorrelated, there is an
anti-correlation. While these types of negative correlations might be attributed to swirls caused
by collective behavior, it is necessary to show an explicit comparison with the uncorrelated
swimmer response because independent swimmers can produce negative correlations due to the change
in hydrodynamics caused by the wall. Figure~\ref{fig:auto-u} shows such a comparison. First
consider the comparison between curves in the plane $x_3 = 0.8H$. We see that the full simulation
correlations are similar to the independent swimmer calculation. The difference between the curves
in slightly larger in the plane $x_3 = 0$. Recall that the independent swimmers are all placed at
the walls, which produce fluid correlations in the center of the channel. The full simulations
however also have some swimmers in the center of the channel. This additional contribution from
the swimmers in the center may be sufficient to account for the difference.

\begin{figure}
\begin{center}
\includegraphics[width=0.47\textwidth]{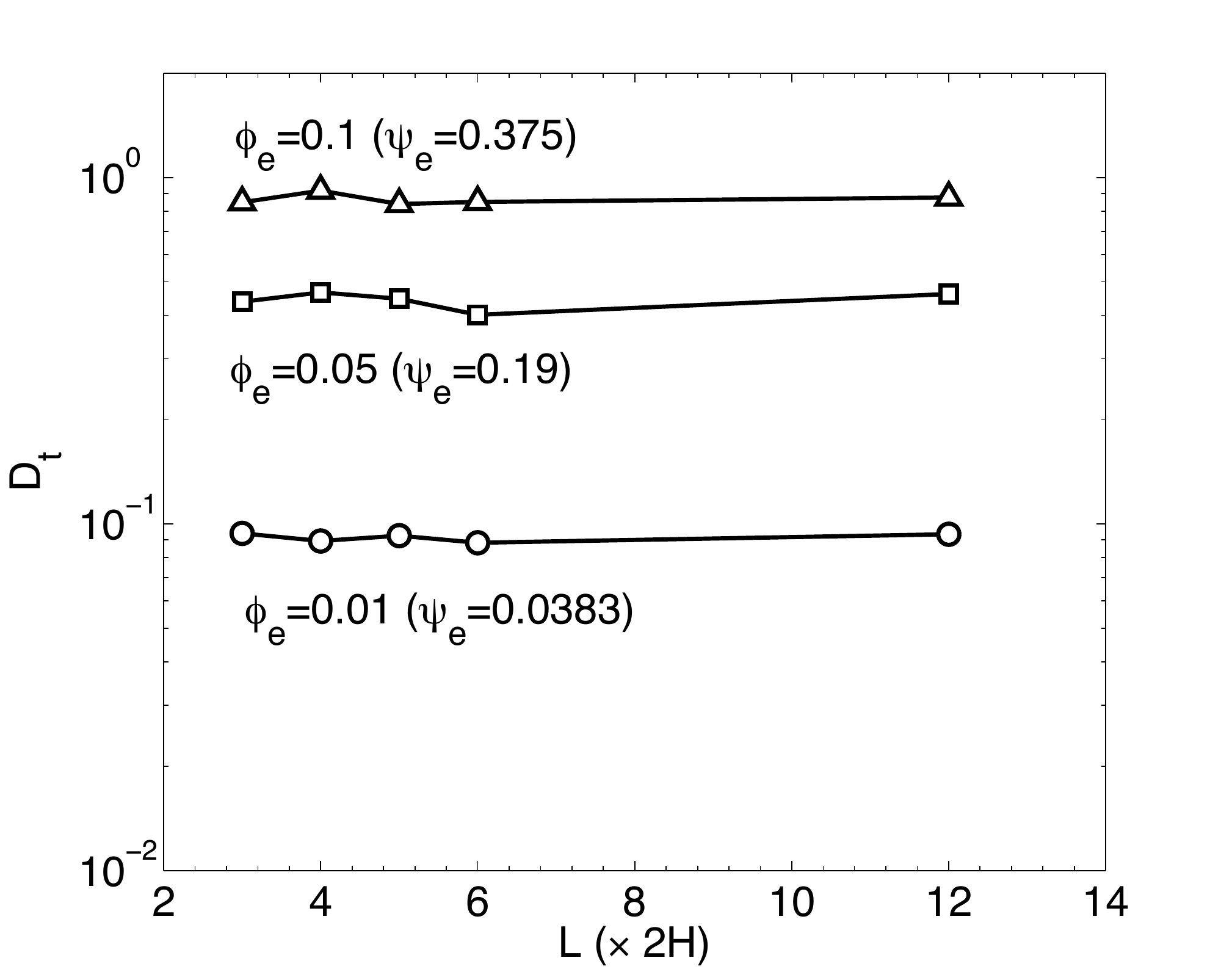}
\caption{System size dependence of the tracer (fluid) diffusion coefficient at three effective
volume fractions. } \label{fig:d-size}
\end{center}
\end{figure}

We now return to the issue of system size effects and our choice of $L$ in the periodic
directions. In order to explore the system size effects on the results, the confinement of $2H=5$
was kept constant while the periodicity of the box was increased from $L=3\times 2H$ to $L=
4\times2H, 5\times2H,6\times2H$ and $12\times 2H$. Underhill {\it et al.}~\cite{underhill-spp}
found that the tracer diffusivity in a three-dimensional periodic unconfined system has a
dependence on system size. Because the tracer diffusivity in the unconfined system had a stronger
system size dependence than the swimmer diffusivity and the velocities, we show in
figure~\ref{fig:d-size} the system size dependence of the tracer diffusion coefficient in the
confined domain. No system size dependence is observed. We also did not observe any system size
dependence of the swimmer diffusivity or swimmer and tracer velocities. This explicitly justifies
our use of $L = 3 \times 2H$ in the data presented earlier.

\begin{figure}
\begin{center}
\includegraphics[width=0.47\textwidth]{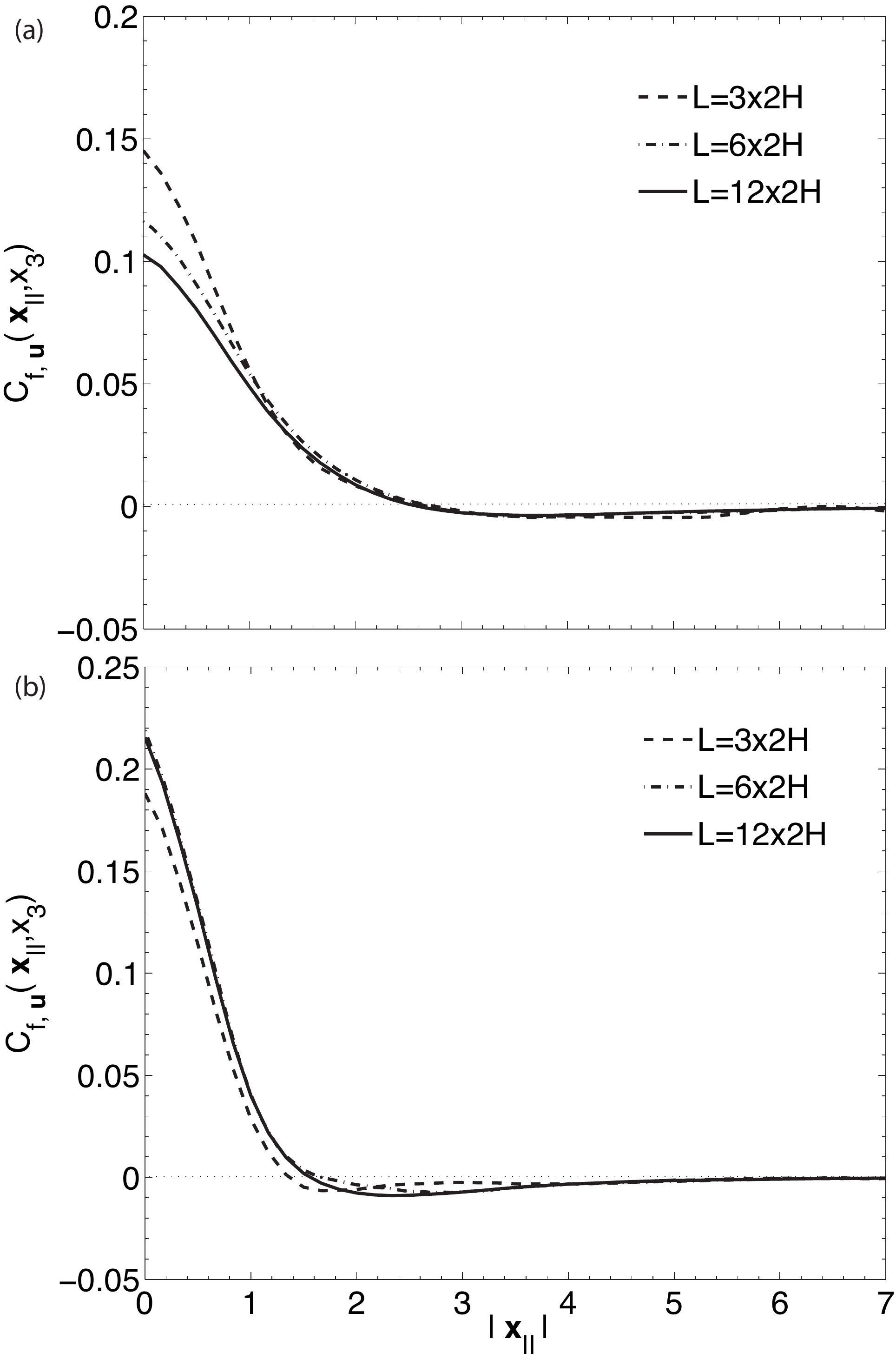}
\caption{Fluid velocity autocorrelation function for $2H=5$ at $\phi_{\mathrm{e}}=0.05$
($\psi_{\mathrm{e}}=0.19$) as a function of the system size. (a) $x_3=0$ (b) $x_3=0.8H$
($x_3=H-1/2$) } \label{fig:auto-u-size}
\end{center}
\end{figure}

Finally, figure~\ref{fig:auto-u-size} illustrates the velocity autocorrelation at $2H=5$ for three
system sizes $L=3\times 2H, 6\times 2H$ and $12\times 2H$. The curves have a slight change with
system size at small separations, though they seem to converge quickly with increasing system
size. Clearly, the lack of changes with a system size in the confined system compared to the
unconfined system is due to the introduction of an additional length scale once the system is
confined. This is not surprising because the confinement restricts flow structures larger than the
confinement and modifies the hydrodynamic interactions between swimmers. If the collective
behavior seen in the unconfined simulations is due to the long-range nature of the hydrodynamic
interactions, screening of these interactions by the walls may affect the collective structures
formed. The effect of hydrodynamic screening on the dynamics of polymers has been studied in
detail~\cite{Jendrejack03b,ChenSlitDiffusion,DoyleSlitScreeningMacro06}. In the polymer
literature, it is the rate of decay of the fluid disturbance with distance (as well as the
rotational symmetry) that determines if screening is present or not. At large enough distance from
a swimmer in the confined domain, the velocity disturbance decays as $r^{-3}$. It is not clear if
this decay is responsible for the results seen here -- that the fluid structures do not depend on
system size provided the periodicity is larger than the scale of confinement.

\section{Conclusions}\label{sec:conclu}
Simulations and theoretical analysis were used to study the dynamics of confined suspensions of
self-propelled particles. The swimmers interact with one another via an excluded volume potential
and hydrodynamic interactions through the fluid. These hydrodynamic interactions are altered by
the confining walls.

We developed a simple theory of the motion of swimmers and tracers which captures the scalings of
the diffusivity and velocities from the simulations. The theory assumes that in the dilute limit
the swimmers form layers near the walls and execute a two-dimensional random walk within the
layer. The theory also shows that even independent swimmers, with no collective behavior, can
produce spatial fluid correlations in a confined domain that do not appear in an unconfined
domain. This is particularly important because swirls in the fluid and negative correlations have
been cited in previous studies as evidence of collective behavior. Using this theory, we were able
to better understand how the scale of these negative correlations in the simulations change with
the degree of confinement. In particular, the negative correlations in the center of the slit are
governed by the separation of the walls, while the negative correlations near the walls are
governed by the size of a swimmer and the separation of the swimmer from the wall. This is
important for understanding experimental observations of swimming suspensions because many
experiments are performed in the presence of walls.

\ack We gratefully acknowledge support from NSF grants CTS-0522386 and DMR-0425880 (Nanoscale
Science and Engineering Center).

\section*{References}
\bibliographystyle{unsrt}
\bibliography{spp_revision090808}

\end{document}